%% file: gravwaves.tex
\documentclass[nofootinbib,a4paper,10pt,eqsecnum,showkeys,showpacs,aps,prd,superscriptaddress,reprint]{revtex4-1}

\usepackage{times}
\usepackage{amsfonts}
\usepackage{amsmath}
\usepackage{amssymb}
\usepackage{natbib}
\usepackage{graphicx}
\usepackage{enumitem}
\usepackage{afterpage}
\usepackage{nccmath}

\input{defs}

\usepackage[breaklinks,colorlinks]{hyperref}
\usepackage{hyperref}
\hypersetup{%
    ,urlcolor=blue
    ,citecolor=blue
    ,linkcolor=blue
    }

\def\apjl{ApJL}
\def\jcap{JCAP}
\def\nat{Nature}

\begin{document}

\title{Gravitational wave constraints on dark sector models}
\author{Richard A. Battye}
\email{richard.battye@manchester.ac.uk}
\affiliation{Jodrell Bank Centre for Astrophysics, School of Physics and Astronomy, The University of Manchester, 
Manchester M13 9PL, U.K.}
\author{Francesco Pace}
\email{francesco.pace@manchester.ac.uk}
\affiliation{Jodrell Bank Centre for Astrophysics, School of Physics and Astronomy, The University of Manchester, 
Manchester M13 9PL, U.K.}
\author{Damien Trinh}
\email{damien.trinh@manchester.ac.uk}
\affiliation{Jodrell Bank Centre for Astrophysics, School of Physics and Astronomy, The University of Manchester, 
Manchester M13 9PL, U.K.}

\label{firstpage}

\date{\today}

\begin{abstract} 
We explore the constraints on dark sector models imposed by the recent observation of coincident gravitational waves 
and gamma rays from a binary neutron star merger, GW170817. Rather than focusing on specific models as has been 
considered by other authors, we explore this in the context of the equation of state approach of which the specific 
models are special cases. After confirming the strong constraints found by others for Horndeski, Einstein-Aether and 
massive gravity models, we discuss how it is possible to construct models which might evade the constraints from 
GW170817 but still leading to cosmologically interesting modifications to gravity. Possible examples are ``miracle 
cancellations" such as in $f(R)$ models, nonlocal models and higher-order derivatives. The latter two rely on the 
dimensionless ratio of the wave number of the observed gravitational waves to the Hubble expansion rate being very 
large 
($\sim10^{19}$) which is used to suppress modifications to the speed of gravitational waves.
\end{abstract}



\maketitle

\section{Introduction}
The detection of gravitational waves from a source almost coincident with a gamma ray burst suggests that the two come 
from the merger of a binary neutron star system \cite{LigoVirgo2017,LigoVirgoIntegral2017}. The measured time 
difference between the two is $\Delta t_{\rm obs}=(1.75\pm 0.05)\,{\rm sec}$ and the distance inferred to the source 
is $D=40^{+8}_{-14}\,{\rm Mpc}$ \cite{LigoVirgo2017a}. 
The difference between two waves emitted a time $\Delta t_{\rm emit}$ apart is given by
\bea
 \Delta t & = & \Delta t_{\rm obs}-\Delta t_{\rm emit} = \frac{D}{c_{\rm G}}-\frac{D}{c_{\gamma}} \nonumber\\
          & = & \frac{D}{c_{\gamma}}\left[\left(1+\frac{\Delta c}{c_{\gamma}}\right)^{-1}-1\right]\,,
\eea
where $c_{\gamma}$ and $c_{\rm GW}$ are the propagation speeds of the photons and gravitational waves respectively, 
and $\Delta c = c_{\rm GW}-c_{\gamma}$. By making the assumptions that $-10<\Delta t_{\rm emit}/{\rm sec}<0$ and 
$\Delta c/c_{\gamma}\ll 1$, and also conservatively using the lower bound on the distance, $D\approx 26\,{\rm Mpc}$, 
one obtains a very strong constraint on the difference between the speed of propagation of gravitational waves and 
photons
\be
 -3\times 10^{-15}<\frac{\Delta c}{c_{\gamma}}<7\times 10^{-16}\,.
\ee
One might question this constraint in that the precise numbers depend very strongly on the unknown 
$\Delta t_{\rm emit}$. However, any value for which one might imagine that it was possible to make a definite 
association between the gravitational wave signal and the counterpart photons still leads to a very strong constraint 
on $\Delta c/c_{\gamma}$ due to the large distance over which the signals have propagated. For example, if 
$|\Delta t_{\rm emit}|<1\,{\rm day}$ then $|\Delta c/c_{\gamma}|<10^{-9}$ which is already a very stringent limit. 
Similar bounds are obtained by the lack of gravitational Cherenkov 
radiation~\citep{Caves1980,Moore2001,Elliott2005,Kimura2012}.

A number of authors \cite{Creminelli2017,Sakstein2017,Baker2017,Ezquiaga2017,Amendola2018,Crisostomi2018} have 
pointed out that this constraint has very severe implications for many, but not all, modified gravity models 
considered in the literature as possible origins of the cosmic acceleration\footnote{We note that the constraining 
power of a simultaneous detection of gravitational waves and electromagnetic counterpart was anticipated by 
\citep{Nishizawa2014,Lombriser2015,Lombriser2016a,Nishizawa2016,Lombriser2017}. In particular, it was shown that large 
scale structure observations would not be able to unequivocally distinguish Horndeski models from the $\Lambda$CDM 
model, but that gravitational waves could break what the authors call ``dark degeneracy" 
\citep{Lombriser2015,Lombriser2016a}.}. The focus of these discussions is mainly on the generalised scalar-tensor (ST) 
models known as Horndeski and beyond Horndeski theories, although there is also some discussion on vector-tensor (VT), 
massive gravity and Ho\v{r}ava models. If these works are to be taken at face value they appear to rule out all but 
the simplest - and observationally least interesting - modified gravity models, implying that observational programmes 
aimed at constraining them using cosmological observations might be wasting their time and significant amounts of 
taxpayer funding.

In our contribution to this discussion\footnote{From now on we will use natural units where 
$c_{\gamma}=\hbar=k_{\rm B}=1$.} 
we do not question the specific calculations presented in these earlier works. However, we do note that dark sector 
models are designed to modify gravity on scales $\sim H_0^{-1}$ whereas the scales relevant to the observations of the 
binary neutron star merger GW170817 are $\sim 10^{14}\, {\rm sec} \sim 10^{-4}H_0^{-1}$ (the lookback time inferred 
from the distance) and $\sim 10^{16}\,{\rm Mpc}^{-1}\sim 10^{19}H_0$ (the wave number computed from the frequency of 
gravitational waves detected). This means that in the context of gravitational wave sources, such as GW170817, there is 
a large dimensionless number ${\rm K}_{\rm grav}=k_{\rm grav}/H_0$ which in principle might be used to suppress 
modifications to gravity on small scales, but which can be very different on large scales. In its very simplest terms 
our argument is that the very wide range of scales between those probed by cosmology and those relevant for the 
detection of gravitational waves means that there is significant room for the construction of models that avoid these 
constraints. In what follows we flesh out our arguments within the equation of state approach to cosmological 
perturbations in dark sector models\footnote{We use the term dark sector to refer to whatever causes cosmic 
acceleration encompassing both dark energy and modified gravity models.}.

\section{Constraining the equation of state approach}
The equation of state approach~\cite{Battye2013,Battye2014} is a phenomenological idea for describing perturbations in 
dark sector models whereby whatever is causing the cosmic acceleration is modelled as an isotropic fluid with equation 
of state $P_{\rm ds}=w_{\rm ds}\rho_{\rm ds}$ where $P_{\rm ds}$ and $\rho_{\rm ds}$ are the pressure and the density 
of the dark sector fluid at the background level, respectively, and $w_{\rm ds}$ is not necessarily a constant, 
but is often considered to be so within the context of present observations. Such a description is sufficient for 
describing observations that are only sensitive to the expansion rate of the Universe. If one wants to also include 
observations sensitive to perturbations, such as those for the cosmic microwave background (CMB) or cosmic shear, then 
it is necessary to also provide an equation of state for the perturbations which encodes how the dark sector 
perturbations respond, allowing the linearised conservation equations for the dark sector fluid to become closed and, 
hence, be solved using standard codes (see, for example, the discussions presented in \cite{Battye2015,Soergel2015}).

Most work to date has focused on the scalar perturbations since they are most relevant to cosmological observations, 
but it can be adapted to the tensor (gravitational wave) sector and indeed the simplicity of the idea is even more 
clear there due to the reduced number of degrees of freedom. Assuming that the $+$ and $\times$ modes of gravitational 
wave evolve identically -which need not be the case- the equation for the evolution of the transverse-traceless 
component of the metric in an FRW universe with a dark sector producing cosmic acceleration is given by
\be\label{eqn:evolution}
 \ddot{h} + 3H\dot{h}+\frac{k^2}{a^2}h = 16\pi G_{\rm N}P_{\rm ds}\Pi_{\rm ds}^{\rm T}\,,
\ee
where $H=\dot{a}/a$ is the Hubble parameter, $G_{\rm N}$ is Newton's constant and $\Pi_{\rm ds}^{\rm T}$ is the tensor 
component of the anisotropic stress. In general, $\Pi_{\rm ds}^{\rm T}\equiv 0$ does not need to imply 
$\Pi_{\rm ds}^{\rm S}\equiv0$ and this could be seen as a simple way to avoid all constraints from GW170817. We would, 
however, see $\Pi_{\rm ds}^{\rm T}\equiv 0$ and $\Pi_{\rm ds}^{\rm S}\ne 0$ as being a little unnatural, but not 
necessarily impossible. In order to solve this equation it is necessary to specify $\Pi_{\rm ds}^{\rm T}$ and by 
similar arguments to those applied to the scalar sector we can write
\be\label{eqn:pit}
 8\pi G_{\rm N}P_{\rm ds}\Pi_{\rm ds}^{\rm T} = C_{\ddot{h}}\ddot{h} + C_{\dot{h}}H\dot{h} + C_{h}H^2h\,,
\ee
where $C_{\ddot h}$, $C_{\dot h}$ and $C_{h}$ are all functions of $a$ and $k$.

When discussing constraints imposed by GW170817, one needs to solve (\ref{eqn:evolution}) inserting the expression in 
(\ref{eqn:pit}) which leads to
\be\label{eqn:evolutionC}
 \ddot{h} + \frac{3-2C_{\dot{h}}}{1-2C_{\ddot{h}}}H\dot{h} + \frac{{\rm K}^2-2C_{h}}{1-2C_{\ddot{h}}}H^2h = 0\,,
\ee
where ${\rm K}=k/(aH)$. In what follows it is more convenient to rewrite (\ref{eqn:evolutionC}) in a simpler form
\be\label{eqn:evolution_beta}
 \ddot{h} + [3+\beta_{\rm M}(a,{\rm K})]H\dot{h}+\beta_{\rm T}(a,{\rm K})H^2h = 0\,,
\ee
where in general the dimensionless coefficients $\beta_{\rm M}$ and $\beta_{\rm T}$ can be a function of both time and 
scale and are related to $C_i$ where $i=\ddot{h}, \dot{h}, h$, via
\be
 \beta_{\rm M} = \frac{2(3C_{\ddot{h}}-C_{\dot{h}})}{1-2C_{\ddot{h}}}\,, \quad 
 \beta_{\rm T} = \frac{{\rm K}^2-2C_{h}}{1-2C_{\ddot{h}}}\,.
\label{eom}
\ee
Specific models for the dark sector predict different forms for the coefficients $\beta_{\rm M}(a,{\rm K})$ and 
$\beta_{\rm T}(a,{\rm K})$ and those already in the literature are presented in Appendix~{\ref{sect:models}} and we 
note that for $\beta_{\rm M}=0$ and $\beta_{\rm T}={\rm K}^2$ we recover the standard general relativistic result. 

The specific choice typically assumed, is that of the Horndeski class of models\footnote{We have shown in 
Appendix~{\ref{sect:models}} that generalized Einstein-Aether models fall into this category, but that massive gravity 
and elastic dark energy models do not.} 
which leads to the specific forms $\beta_{\rm M}=\alpha_{\rm M}(a)$ and $\beta_{\rm T}=[1+\alpha_{\rm T}(a)]{\rm K}^2$, 
and it is this specific choice that leads to the very strong conclusions reported in 
\cite{Creminelli2017,Sakstein2017,Baker2017,Ezquiaga2017,Amendola2018,Crisostomi2018}, for example. 
In particular it has been argued that the constraints from GW170817 imply that $|\alpha_{\rm T}|<10^{-15}$ and hence 
that it is reasonable to assume that $\alpha_{\rm T}\equiv 0$ in these models. What we have argued here is that this 
specific form could be too restrictive and in particular there is room for the speed of gravitational waves being 
dependent on ${\rm K}$. 
In Sec.~\ref{sec:form} we will investigate how it might be possible to avoid these conclusions.

Before doing this we will address the solution of (\ref{eqn:evolution_beta}) using the Wentzel-Kramers-Brillouin (WKB) 
approximation as recently done in detail, including source terms, in~\cite{Nishizawa2017,Arai2017}, to which we refer 
for more details and for a more general discussion. Assuming a solution of the form $h=A(t)\exp[\imath\psi(t)]$, 
Eq.~(\ref{eqn:evolution_beta}) is equivalent to the following two sets of equations:
\begin{eqnarray}
 \ddot{A} - A\dot{\psi}^2 + (3+\beta_{\rm M})H\dot{A} + \beta_{\rm T}H^2A & = & 0\,,\\
 2\dot{A}\dot{\psi} + A\ddot{\psi} + (3+\beta_{\rm M})HA\dot{\psi} & = & 0\,,
\end{eqnarray}
where the two equations are derived from the real and imaginary parts, respectively. The condition we impose is that 
the amplitude of the gravitational wave is slowly changing with respect to the frequency of the wave itself $\psi$, 
therefore it is reasonable to assume that $\dot{\psi}^2\gg\ddot{A}/A$ and $\dot{\psi}^2\gg(3+\beta_{\rm M})H\dot{A}/A$ 
which is equivalent to the oscillation timescale being much faster than the Hubble rate. This would be true for 
gravitational waves from GW170817, and similar objects, but is not necessarily relevant on cosmological scales. Under 
these conditions, the first equation reduces to $\dot{\psi}=\sqrt{\beta_{\rm T}}H$ whose solution is
\be
 \psi = \int_a\sqrt{\beta_{\rm T}}\frac{da^{\prime}}{a^{\prime}}\,,
\ee
and the second one to $\partial_t\ln{(A^2\dot{\psi})}=-(3+\beta_{\rm M})H$ whose solution is
\be
 A = \frac{\exp{\left[-\frac{1}{2}\int_a(3+\beta_{\rm M})\frac{da^{\prime}}{a^{\prime}}\right]}}
          {(\sqrt{\beta_{\rm T}}H)^{1/2}}\,.
\ee
The full WKB solution is
\begin{fleqn}
 \bea
  h({\rm K},t) = && \frac{h_0}{(\sqrt{\beta_{\rm T}}H)^{1/2}}
                    \exp{\left[-\frac{1}{2}\int_{a_i}^{a(t)}(3+\beta_{\rm M})\frac{da^{\prime}}{a^{\prime}}\right]}
                    \nonumber\\
                 && \times
                    \exp{\left[\imath\int_{a_i}^{a(t)}\sqrt{\beta_{\rm T}}\frac{da^{\prime}}{a^{\prime}}\right]}\,,
 \eea
\end{fleqn}
where $h_0$ represents the amplitude of the wave at $a=a_i=a(t_i)$.

We now evaluate the dispersion relation for gravitational waves and derive expressions for the phase 
$v_{\rm p}=\omega/k$ and the group velocity $v_{\rm g}=d\omega/dk$. The frequency is 
$\omega({\rm K}) = \dot{\psi} = \sqrt{\beta_{\rm T}}H$, which leads to
\begin{fleqn}
 \be
  v_{\rm p}(k) = \frac{\sqrt{\beta_{\rm T}}}{a{\rm K}}\,, \quad 
  v_{\rm g}(k) = \frac{\beta_{\rm T}^{\prime}}{2a^2{\rm K}v_{\rm p}}
               = v_{\rm p}\frac{{\rm K}\beta_{\rm T}^{\prime}}{2\beta_{\rm T}}\,,
 \ee
\end{fleqn}
where a prime denotes the derivative with respect to ${\rm K}$. These expressions are very simple and encompass a wide 
range of dark sector models. For a more general discussion on the group velocity of gravitational waves, we refer the 
reader to \cite{Balek2009}, but from the point of view of the present discussion it is important to note two points. 
First, the speed of gravitational waves only depends on $\beta_{\rm T}$ and $\beta_{\rm M}$ is 
unconstrained\footnote{It is possible for the observations of coincident gravitational and electromagnetic waves to be 
used to infer a distance measure and a redshift and hence for the construction of a Hubble diagram based on these 
``standard sirens". Indeed this method has already been used to infer a measurement of the Hubble 
constant~\cite{LigoVirgo2017a}. In future it might be possible to use this approach to infer constraints on 
$\beta_{\rm M}$~\cite{Lombriser2016a,Amendola2017b}.}. 
In addition it is clear that, for a general dependence of $\beta_{\rm T}$ on ${\rm K}$, $v_{\rm p}\ne v_{\rm g}$. The 
observations of coincident electromagnetic and gravitational waves refer to the coincidence of detection of energy and 
hence refer specifically to the group velocity and not to the phase velocity. This distinction is not relevant in the 
Horndeski case where $v_{\rm p}=v_{\rm g}$, but we need to be slightly more careful here.

All the models discussed in the Appendix can be parameterized by the form 
$\beta_{\rm T}=(1+\alpha_{\rm T}){\rm K}^2+{\rm M}_{\rm GW}^2$, where $\alpha_{\rm T}$ is a function of time and 
${\rm M}_{\rm GW}=m_{\rm GW}/H$ is the time dependent, dimensionless graviton mass. In this case the two velocities 
read
\begin{fleqn}
 \be
  v_{\rm p}({\rm K}) = \frac{1}{a}\sqrt{1+\alpha_{\rm T}+\frac{{\rm M}_{\rm GW}^2}{{\rm K}^2}}\,, \; \; 
  v_{\rm g}({\rm K}) = \frac{1+\alpha_{\rm T}}{a^2v_{\rm p}}\,.
 \ee
\end{fleqn}
If ${\rm M}_{\rm GW}\ll {\rm K}_{\rm grav}\sim 10^{19}$, which one would naturally expect, then we have that 
$v_{\rm p}=v_{\rm g}=\frac{1}{a}\sqrt{1+\alpha_{\rm T}}$ and hence we derive the constraint $|\alpha_{\rm T}|<10^{-15}$ 
as previously deduced. However, we see that there is no extra constraint imposed by GW170817 on ${\rm M}_{\rm GW}$. 
This is due to the suppression of this quantity by the large dimensionless number ${\rm K}_{\rm grav}$. However, the 
massive graviton could still have significant cosmological effects. In the next section we will attempt to develop this 
line of argument to more general dark sector models.

\section{Form of the equation of state for the dark sector}
\label{sec:form}
In the previous section we have argued that the evolution of cosmological gravitational waves in the most general dark 
sector models can be parameterized by $\beta_{\rm M}\equiv\beta_{\rm M}(a,{\rm K})$ and 
$\beta_{\rm T}=\beta_{\rm T}(a,{\rm K})$ and the very specific case of $\beta_{\rm M}\equiv\alpha_{\rm M}(a)$ and 
$\beta_{\rm T}\equiv[1+\alpha_{\rm T}(a)]{\rm K}^2$ assumed by most authors leads to very strong constraints from 
GW170817. In this section we will explore how it might be possible to evade these constraints in more general models.

Before this we should make an important point concerning our choice to parameterise these functions in terms of 
the dimensionless combination ${\rm K}=k/(aH)$ which is $\gg 1$ in the regime relevant to gravitational waves from 
GW170817. All dark sector models could be considered to be unnatural in some way since the timescale of the age of 
Universe, $H_0^{-1}$, has been introduced to them by hand. This is manifest even in models with a cosmological constant 
where $\Lambda\propto H_0^{2}$ - this is often known as the timescale problem or 
``why is $\Omega_{\Lambda}\sim \Omega_{\rm m}$ today?" We do not attempt to solve this problem, but our argument is 
that once one accepts the addition of this new dimensionful quantity into the problem, one is not further increasing 
the complexity by reusing it. The significant consequence of this is that it is natural for cosmological 
observations to probe in the regime ${\rm K}\ll 1$, while the solar system, where there are very stringent constraints 
on the nature of gravitational interactions~\cite{Bertotti2003,Will2006}, and GW170817 are in the regime 
${\rm K}\gg 1$. Hence, the constraints imposed by GW170817, while extremely strong in the regime of validity, only 
impose constraints in a regime very different to that probed by cosmological observations and hence one does not have 
to work too hard to construct a dark sector model capable of explaining large-scale cosmic acceleration while still 
being compatible with measurements on smaller scales.

In order to understand how one might avoid the constraints imposed by GW170817, let us consider the case where the 
dispersion relation is parameterized by some function $\chi({\rm K})$ defined by 
\be\label{eqn:dr}
 \omega^2={\rm K}^2H^2\left[1+\chi({\rm K})\right]\,,
\ee
in which case the coefficients of the equation of state can be written as 
\be
 C_{h}={\rm K}^2\left[C_{\ddot{h}}+\chi\left(C_{\ddot{h}}-\frac{1}{2}\right)\right]\,.
\ee
With this form for the dispersion relation, $\beta_{\rm T}={\rm K}^2\left[1+\chi({\rm K})\right]$ and
\begin{fleqn}
 \be
  v_{\rm p}=\frac{1}{a}\sqrt{1+\chi}\,, \quad 
  v_{\rm g}=\frac{1}{a}\left[\sqrt{1+\chi}+\frac{{\rm K}\chi^{\prime}}{2\sqrt{1+\chi}}\right]\,.
 \ee
\end{fleqn}
For the case below ensuring that $v_{\rm g}\approx 1/a$, which is what the observations require, is equivalent to 
$v_{\rm p} \approx 1/a$ and therefore we will concentrate on the simpler case of ensuring $v\approx 1/a$.

The form of $\chi({\rm K})$ in the regime ${\rm K}\gg 1$ governs the evolution of gravitational waves in the 
regime relevant to GW170817. If spatial derivatives enter in second order combinations (for example, $(\nabla_i F)^2$, 
$\nabla_i\nabla_j F$ for some scalar function $F$) then it seems reasonable to expand $\chi({\rm K})$ as a power series 
in ${\rm K}^2$. The observed properties of gravitational waves suggest that terms with positive powers of ${\rm K}^2$ 
are excluded and therefore we consider\footnote{This choice, written as a  power series, appears to diverge as 
${\rm K}\rightarrow 0$. It is necessary the actual function which this power series represents would have a finite 
limit and is regularised in some way in order to avoid extreme behaviour in the infrared regime of the theory. Such 
behaviour would lead to a violation of causality. Simple function which has this property is 
$\chi({\rm K})\propto ({\rm K}_0^2+{\rm K}^2)^{-1}$ for some constant ${\rm K}_0$.}
\be\label{eqn:chi_K}
 \chi({\rm K})=\sum_{n=0}^{\infty}\frac{\chi_n}{{\rm K}^{2n}}\,,
\ee
where the dimensionless coefficients $\chi_{n}\equiv\chi_n(a)$ are chosen so that $1+\chi$ remains $>0$ for all 
${\rm K}$. The first two coefficients have physical interpretations: $\chi_0=\Delta c/c_{\gamma}$ is the modification 
to the speed of propagation of gravitational waves constrained to be $|\chi_0|\ll 10^{-15}$ and 
$\chi_1={\rm M}^2_{\rm GW}=m^2_{\rm GW}/H^2$ is the dimensionless mass associated with a graviton mass $m_{\rm GW}$. 
Observations of the gravitational waves event GW150914 lead to a relatively weak limit of 
$m_{\rm GW}\le 1.2\times 10^{-22}\,{\rm eV}$ which implies that 
${\rm M}_{\rm GW}(a=1) \lesssim 10^{10}$~\citep{LigoVirgo2016}. 
We note that there is a stronger constraint of 
$m_{\rm GW}\lesssim 10^{-30}\,{\rm eV}$, $M_{\rm GW}(a=1) \lesssim 10^{3}$ enforced by consideration of gravity in the 
solar system \citep{deRham2017} and from weak lensing data \citep{Choudhury2004,Desai2018,Rana2018}.

In order to investigate possible models that might be able to avoid constraints from GW170817 it is interesting to 
consider some special cases.

\begin{itemize}
\item The simplest possible case is where $\chi\equiv 0$ which implies that $C_{h}={\rm K}^2C_{\ddot{h}}$. An example 
of such a model is the $f(R)$ gravity model, or indeed any Horndeski model with $\alpha_{\rm T}\equiv 0$. We describe 
models with this property as having a ``miracle cancellation,'' in that they have $\alpha_{\rm T}\equiv 0$ without 
having $\Pi_{\rm ds}^{\rm T}=0$ and more importantly from the point of view of having interesting observational 
signatures due to the evolution of dark sector perturbations. In fact all Horndeski models with 
$G_4\equiv G_4(\phi)$ and $G_5$ constant lead to such miracle cancellations. The conditions required for these 
miracle cancellations in generic scalar-tensor theories were determined in \cite{Bettoni2017}.

\item If $C_{\ddot{h}}$ is independent of ${\rm K}$ and consider the possibility of 
$\chi=\chi_1/{\rm K}^2+\chi_2/{\rm K}^4$ as the simplest case which gives something beyond the graviton mass, then 
$C_{h}=B_2{\rm K}^2+B_0+B_{-2}/{\rm K}^2$ for some coefficients $B_2$, $B_0$ and $B_{-2}$ which are functions of the 
scale factor. In order to construct such a model with negative powers of ${\rm K}$ it may be necessary to introduce 
nonlocal modifications to gravity so that the equation of state contains terms such as $\nabla^{-2}h$. To see this 
more explicitly, let us consider for simplicity a model where the graviton mass is zero and the only term in the 
series expansion is $\chi_1$. The equation of motion for the transverse-traceless degrees of freedom $h$ is
\bea\label{eqn:NL_GW}
 \ddot{h} + 3H\dot{h} + {\rm K}^2H^2h + \frac{\chi_1}{{\rm K}^2} = 0 &&\nonumber\\
 \leftrightarrow
 \ddot{h} + 3H\dot{h} - \frac{1}{a^2}\nabla^2 h + \int d^3{\bf x}^{\prime}K({\bf x}-{\bf x}^{\prime})
 h({\bf x}^{\prime},t) = 0&&,\nonumber\\
\eea
where $K({\bf x}) = \chi_1 |{\bf x}|^{-1}$ would give rise to such a behaviour and other suitably regularised kernels 
could be computed to achieve other limiting behaviours for ${\rm K}\gg 1$ (i.e. higher order inverse powers of 
${\rm K}$). Constructing a Lagrangian which leads to this kind of evolution for the gravitational waves may be quite 
challenging, but it is not obviously impossible. 
We note that a model containing a nonlocal ``mass" term $\propto R\,\Box^{-2} R$ where $R$ is the Ricci scalar has 
been studied in a number of works with the conclusion that the model gives rise to a local equation for the 
traceless-transverse degrees of freedom where gravitational waves propagate at the speed of light 
\citep[for example][]{Maggiore2014,Belgacem2018}. Since this model has a nonvanishing $\Pi^{\rm T}_{\rm ds}$, it 
can be seen as another example, together with $f(R)$ models, of miracle cancellation. 
This happens because the model can be recast into a multiscalar-tensor theory. We feel though, this issue warrants 
further investigation since it has interesting cosmological consequences while at the same time surviving the 
constraints of GW170817.

In general it is not known how to build a nonlocal Lagrangian that gives rise to an integral term of the form as 
in (\ref{eqn:NL_GW}), but one can follow the approach of \cite{Chicone2013,Mashhoon2014} and enforce it at the level of 
the equations of motion. As shown in these works, this nonlocal term will propagate up to the equations of motion for 
the gravitational waves. Since there is no underlying physical argument which leads to a form for the kernel function 
$K({\bf x}-{\bf x}^{\prime})$, it is necessary to use phenomenological parametrisations, which can, nevertheless, be 
constrained by data, as for example where a nonlocal Poisson equation is valid; we refer to \cite{Mashhoon2014} for 
details of specific models.

\item A more general form for $C_{\ddot{h}}=A_0+A_2{\rm K}^2$ and the same form for $\chi$ as in the last example in 
which case $C_{h}=B_4{\rm K}^4+B_2{\rm K}^2+B_0+B_{-2}/{\rm K}^2$ with $B_4$, $B_2$, $B_0$ and $B_{-2}$ again scale 
factor dependent coefficients. If one were to make the particular choice $A_0=1/2$ then one finds that $B_{-2}\equiv 0$ 
removing, by a specific cancellation, the need for nonlocal inverse powers of ${\rm K}$ and 
$\beta_{\rm T}\approx B_4{\rm K}^2/A_2$ at large ${\rm K}$ so observations require $B_4/A_2=1+{\cal O}(10^{-15})$. In 
this case, it would be necessary for the equation of state to contain terms such as $\nabla^2\ddot{h}$ and $\nabla^4h$. 
This is for example the case for higher-order-derivatives theories (see, for example, \citep{Stelle1978}). 
One might be concerned that such models might suffer from Ostrogradsky ghosts or other instabilities since these often 
appear in theories with higher order derivatives. Of course, one can easily construct models without them, $f(R)$ and 
more general Horndeski models being examples, and by construction - since we have defined a positive definite 
dispersion relation - our suggestions would automatically avoid them.

A simple example of such an equation of motion for the transverse-traceless degrees of freedom is
\bea
 \ddot{h} + \frac{\nabla^2\ddot{h}}{a^2H^2} + 3H\dot{h} + 3\frac{\nabla^2\dot{h}}{a^2H} - \frac{\nabla^2h}{a^2} + 
 \frac{\nabla^4h}{a^4H^2} + m^2_{\rm GW}h = 0 && \nonumber\\
 \leftrightarrow
 \ddot{h} + 3H\dot{h} + \frac{M^2_{\rm GW}+{\rm K}^2+{\rm K}^4}{1-{\rm K}^2}H^2h = 0\,,&&\nonumber\\
\eea
where we have specifically chosen the functional form of $C_{\dot{h}}$ to recover the standard friction term of 
general relativity, which need not be the case. Finally, $\nabla^4$ represents the biharmonic 
operator\footnote{In three dimensions, we have 
$\nabla^4=\tfrac{\partial^4}{\partial x^4}+\tfrac{\partial^4}{\partial y^4}+\tfrac{\partial^4}{\partial z^4}+
2\tfrac{\partial^4}{\partial x^2\partial y^2}+2\tfrac{\partial^4}{\partial x^2\partial y^2}+
2\tfrac{\partial^4}{\partial y^2\partial z^2}$.}.

\end{itemize}

\noindent We note that this list of possibilities is far from exhaustive and indeed the details of the last two 
depend quite strongly on the choice of $\chi$. Nonetheless we believe that one would come to similar qualitative 
conclusions in more general cases.

We note that an approach very similar to ours has been suggested by \cite{Mirshekari2012,Yunes2016,Samajdar2017} 
to take into account quantum-mechanical effects which predict a small amount of violation to the otherwise accepted 
Lorentz covariance of physical laws. In this approach, the modified dispersion relation is defined by
\be\label{eqn:LV}
 E^2 = p^2 + m^2_{\rm GW} + \mathbb{A}p^{\alpha}\,,
\ee
where $\mathbb{A}$ defines the magnitude of the deviations from the standard picture (with units [energy]$^{2-\alpha}$) 
and $\alpha$ is a dimensionless constant. The models become particularly appealing for $\alpha<2$ as they provide a 
screening length. In addition to the Compton length $\lambda_{\rm GW}=1/m_{\rm GW}$ associated to the graviton mass, 
there is a characteristic scale $\lambda_{\mathbb{A}}=\mathbb{A}^{1/(\alpha-2)}$ associated with Lorentz violation 
\citep{Samajdar2017}. Despite being phenomenological, the parameterized form of the dispersion relation in 
(\ref{eqn:LV}) can accommodate some particular classes of models, as described in \cite{Mirshekari2012} and 
\citep{Yunes2016}.

Let us now rewrite the dispersion relation in a form more suitable for the goals of this work. Upon the following 
identifications, $E=\omega$ and $p=k$, we obtain
\be\label{eqn:omega_LV}
 \omega^2 = {\rm K}^2H^2\left(1 + \frac{{\rm M}_{\rm GW}^2}{{\rm K}^2} + 
            \frac{\mathbb{A}}{(H{\rm K})^{2-\alpha}}\right)\,,
\ee
and from (\ref{eqn:dr}), assuming $\alpha=-2$, we can read off 
$\chi({\rm K})= {\rm M}_{\rm GW}^2/{\rm K^2} + \mathbb{A}/(H{\rm K})^4$. We can easily see that 
$\chi_1 = {\rm M}_{\rm GW}^2$ and $\chi_2 = (\lambda_{\mathbb{A}}/H)^4$ is the term arising from Lorentz violation.

\section{Conclusions}

In this paper we have attempted to address the question of whether it is possible to construct dark sector models which 
can naturally evade the very strong constraints imposed by GW170817 while still giving rise to cosmologically 
interesting signatures. Within the Horndeski class of scalar tensor models usually considered there is a strong 
constraint which restricts the space of models. This restriction prima-facie forces one into the regime where 
$G_{4}\equiv G_4(\phi)$ and $G_5\equiv 0$. Models where $G_4$ is a constant which fall into this class are much less 
observationally interesting since they do not have anisotropic stress and indeed they could be thought of as dark 
energy models, as opposed to a genuine modified gravity model where the cosmic acceleration is a self-acceleration 
effect~\cite{Lombriser2015,Lombriser2016a}. An alternative that avoids this constraint is the introduction of the mass 
for the graviton or an equivalent effect due to elastic dark energy. The generalisation we have advocated is to allow 
the coefficients describing the evolution of cosmological gravitational waves (\ref{eqn:evolution_beta}) to have 
arbitrary dependence on ${\rm K}$ parameterized by $\beta_{\rm M}(a,{\rm K})$ and $\beta_{\rm T}(a,{\rm K})$. The 
specific choice $\beta_{\rm T}={\rm K}^2(1+\alpha_{\rm T})+{\rm M}_{\rm GW}^2$ is the one which is strongly constrained 
as described in previous works and we concur with these conclusions. More generally, observations force 
$\beta_{\rm T}\approx H^2{\rm K}^2$ at ${\rm K}={\rm K}_{\rm grav}\approx 10^{19}$, but say nothing about the larger 
scales relevant to cosmology and, at least at this level of sophistication, it seems perfectly reasonable to imagine a 
simple functional form leading to this kind of behaviour.

The strong constraints on $\Delta c/c_{\gamma}$ come from the large distance between the source of the gravitational 
waves and their detection on earth by LIGO. In order to avoid this constraint we have suggested to use the small 
dimensionless number ${\rm K}_{\rm grav}^{-1}$ to suppress the effects of a modification of gravity that might lead to 
cosmologically interesting effects on the scales relevant to gravitational wave sources. We have only talked about the 
basic idea behind this suppression mechanism. We have not constructed explicit models at the level of a Lagrangian and 
indeed we acknowledge that it might be difficult to achieve in practice. Other than a miracle cancellation similar 
to that found in $f(R)$ models, we identified two possible directions for further exploration: nonlocal models and 
higher-order derivatives, providing concrete examples of equations of motions which lead to such dispersion relations.

One thing that we should point out is that the suppression mechanism used for the gravitational wave sector of the 
theory could also operate in the scalar density perturbation sector and in principle be used to suppress modifications 
to gravity on solar system scales characterised by ${\rm K}_{\rm solar}\approx 4\times 10^{14}$ (corresponding to 
lengthscales $\sim 10\,{\rm au}$). In Appendix~\ref{sect:suppression} we have outlined some of the basics behind this 
idea. In its simplest possible terms, the coefficients $C_{ij}$ from (\ref{scalartheory}) are chosen so that 
modifications to gravity quantified by two of $\mu_{\psi}$, $\mu_{\phi}$, $\eta$ and $\Sigma$ are equal to their 
general relativistic values when ${\rm K}\sim{\rm K}_{\rm solar}$, but the cancellations of coefficients required to 
achieve this would only be true up to inverse powers of ${\rm K}$.

Since we have only outlined the basic ideas, there is clearly much detailed work to be done to develop fully fledged 
theories. Nonetheless we believe we have made a simple argument that one can develop theories which are compatible with 
general relativity on scales $\sim{\rm K}_{\rm grav}$ and ${\rm K}_{\rm solar}$ while being interestingly different for 
${\rm K}\sim 1$. Indeed the fact that ${\rm K}_{\rm grav}>{\rm K}_{\rm solar}$ suggests that it is not at all 
unreasonable to think that a suppression mechanism which works on solar system scales would allow one to avoid the 
constraints from GW170817.

\section*{Acknowledgements}
\noindent R.A.B. and F.P. acknowledge support from STFC grant ST/P000649/1 and useful discussions with Lucas Lombriser, 
Pedro Ferreira, Jens Chluba and Boris Bolliet. D.T. is supported by an STFC studentship. We thank an anonymous 
referee whose comments helped us to improve the scientific content of this work. D.T. also thanks BritGrav2018 for 
useful discussions on Lorentz-violating models.

\appendix

\section{Example equations of state}\label{sect:models}
 
In this Appendix we present a survey of the coefficients $C_{ij}$ for some of the modified gravity models which have 
been already evaluated in literature and show that these results lead to the conclusions that match the results 
found by others.

\medskip
\noindent (1) {\it Horndeski theories}. These are the most general scalar-tensor theories compatible with second-order 
time evolution. They are specified in terms of four free functions $G_i(\phi,X)$ for $i=2,5$ where $\phi$ is the 
scalar field and $X=-\textstyle\frac{1}{2}\nabla_{\mu}\phi\nabla^{\mu}\phi$ is the canonical kinetic term. The 
equation of state for the tensor sector in these models is given by
\bea\label{eqn:PiTHorndeski}
 8\pi G_{\rm N} P_{\rm ds}\Pi^{\rm T}_{\rm ds} = &&
   -\frac{1}{2}\left\{
    \left(\frac{m^2}{m_{\rm pl}^2}-1\right)\ddot{h}\right.\nonumber\\
    && +\left[\frac{m^2}{m_{\rm pl}^2}(3+\alpha_{\rm M})-3\right]H\dot{h}\nonumber\\
    && \left.+\left[\frac{m^2}{m_{\rm pl}^2}\left(1+\alpha_{\rm T}\right)-1\right]{\rm K}^2H^2h\right\}\,,
\eea
and hence we can read off
\bea
 && C_{\ddot{h}} = -\frac{1}{2}\left(\frac{m^2}{m_{\rm pl}^2}-1\right)\,, \quad 
    C_{\dot{h}} = -\frac{1}{2}\left[\frac{m^2}{m_{\rm pl}^2}(3+\alpha_{\rm M})-3\right]\,, \nonumber\\ 
 && C_{h} = -\frac{1}{2}\left[\frac{m^2}{m_{\rm pl}^2}(1+\alpha_{\rm T})-1\right]{\rm K}^2\,,
\eea
where $m$ represents the effective Planck mass which can be, in general, a function of time, $\alpha_{\rm T}$ the 
excess speed of gravitational waves and $m_{\rm pl}=G^{-1/2}_{\rm N}$ the bare Planck mass.
$\alpha_{\rm M}=\tfrac{1}{H}\tfrac{d\ln{m^2}}{dt}$ is the logarithmic time variation of the effective Planck mass. 
These parameters, together with $\alpha_{\rm B}$ and $\alpha_{\rm K}$ (these last two important for the scalar sector) 
completely define Horndeski theories and have been introduced for the first time in \cite{Bellini2014}. The 
identification with the $\beta_i$ functions introduced in (\ref{eqn:evolution_beta}) is now trivial: 
$\beta_{\rm M}=\alpha_{\rm M}(a)$ and $\beta_{\rm T}=[1+\alpha_{\rm T}(a)]{\rm K}^2$ and the observations of GW170817 
imply that $|\alpha_{\rm T}|<10^{-15}$. 

We can express $\alpha_{\rm T}$ in terms of the functions $G_{4}$ and $G_{5}$ as\footnote{Expressions for $m$ and all 
the $\alpha_i$ functions can be found in \cite{Bellini2014,Gleyzes2013,Gleyzes2014,Tsujikawa2015} in terms of the 
$G_i$ functions.}
\be
 \alpha_{\rm T} = \frac{X\left[2G_{4,X}-2G_{5,\phi}-\left(\ddot{\phi}-H\dot{\phi}\right)G_{5,X}\right]}
                       {G_4-2XG_{4,X}+XG_{5,\phi}-\dot{\phi}HXG_{5,X}}\,,
\ee
which reduces to
\be
 \alpha_{\rm T}=\frac{2XG_{4,X}}{G_4}\left(1-\frac{2XG_{4,X}}{G_4}\right)^{-1}\,,
\ee
when $G_5$ is a constant, which is equivalent to setting $G_5\equiv 0$ by integration by parts. From this we can 
deduce that $\alpha_{\rm T}\ll 1$ can be {\rm achieved} when $XG_{4,X}/G_4\ll 1$ (i.e., the slope of $G_4$ with 
respect to $X$ is close to zero). The most natural way to achieve this is when $G_4\equiv G_{4}(\phi)$ although there 
are other possibilities.

There are two interesting and well studied subclasses of the Horndeski model: 
\begin{itemize}
\item Quintessence~\citep{Ford1987,Peebles1988,Ratra1988a,Wetterich1988,Caldwell1998}, 
$k$-essence~\citep{ArmendarizPicon1999,Chiba2000,Hamed2004,Piazza2004,Scherrer2004,Mukhanov2006} and kinetic gravity 
braiding (KGB) models~\citep{Deffayet2010,Pujolas2011} are subclasses of the Horndeski theory with $G_4$ constant and 
$G_5=0$ and hence $C_{\ddot{h}}=C_{\dot{h}}=C_{h}\equiv0$. All of these minimally coupled scalar field models 
predict no modifications to the evolution of gravitational waves and, therefore, survive constraints from GW170817. 
Of course, this should be no surprise since such models have no anisotropic stress at all, but this also implies that 
they only weakly impact on cosmological observables such as the CMB and cosmic shear~\cite{Battye2015}. 

\item $f(R)$ models are also a subclass for which $m^2=m_{\rm pl}^2\left(1+\tfrac{df}{dR}\right)$ and 
$\alpha_{\rm T}=0$ where $f(R)$ is the modification to the Einstein-Hilbert action. In this class of models, 
$C_h={\rm K}^2C_{\ddot{h}}$ which is the miracle cancellation discussed in Sec.~\ref{sec:form} and hence this 
class of models survives the constraints imposed by GW170817 by having $\alpha_{\rm T}\equiv 0$, but 
$\Pi_{\rm ds}^{\rm T}\ne 0$.
\end{itemize}

\medskip
\noindent (2) {\it Generalised Einstein-Aether theories}. Einstein-Aether theories \citep{Zlosnik2008} are 
vector-tensor theories of gravity which involve the addition of a timelike unit normalised vector field $A^{\mu}$, such 
that $A^\mu A_\mu + 1 = 0$, with a Lagrangian described by a generalised function $F({\cal K})$ where
\be
 {\cal K} = \frac{1}{m_{K}^2}{K^{\alpha\beta}}_{\mu\nu}\nabla_{\alpha}A^{\mu}\nabla_{\beta}A^{\nu}\;,
\ee
and the rank-4 tensor is defined as
\be
 {K^{\alpha\beta}}_{\mu\nu} = c_1g^{\alpha\beta}g_{\mu\nu} + c_2\delta^{\alpha}_{\mu}\delta^{\beta}_{\nu} + 
                              c_3\delta^{\alpha}_{\nu}\delta^{\beta}_{\mu} + c_4A^{\alpha}A^{\beta}g_{\mu\nu}\,.
\ee
The $c_i$ are dimensionless constants and $m_{K}$ has dimensions of mass. The timelike unit norm constraint ensures 
only one scalar degree of freedom propagates which makes this theory similar to the scalar-tensor theories discussed 
above. It can be shown that
\be
 8\pi G_{\rm N}P_{\rm ds}\Pi^{\rm T}_{\rm ds} = -\frac{1}{2}c_{13}
 \left[\frac{dF}{d{\cal K}}\ddot{h} + \left(3\frac{dF}{d{\cal K}}H + 
 \frac{d^2F}{d{\cal K}^2}\dot{\cal K}\right)\dot{h}\right]\,,
\ee
where $c_{13}=c_1+c_3$ from which we can read off
\bea
 && C_{\ddot h} = -\frac{1}{2}c_{13}\frac{dF}{d{\cal K}} \,, \quad 
    C_{\dot h} = -\frac{1}{2}c_{13}\left(3\frac{dF}{d{\cal K}}+\frac{d^2F}{d{\cal K}^2}\frac{\dot{\cal K}}{H}\right)
                 \,, \nonumber\\
 && C_{h} = 0 \,.
\eea
In terms of the $\beta_i$ parameters, we find 
$\beta_{\rm M}=\alpha_{\rm M}=\tfrac{1}{H}\tfrac{d\ln{m^2}}{dt}$ with an effective Planck mass 
$m^2=m_{\rm pl}^2\left(1+c_{13}\frac{dF}{d{\cal K}}\right)$ and 
$\beta_{\rm T}={\rm K}^2(1+\alpha_{\rm T})$ where 
$\alpha_{\rm T}=-c_{13}\tfrac{dF}{d{\cal K}}\left(1+c_{13}\tfrac{dF}{d{\cal K}}\right)^{-1}$.

From this we can deduce that $v_{\rm p}=\left(1+c_{13}\tfrac{dF}{d{\cal K}}\right)^{-1/2}$. The tight constraints 
$\Delta c/c_{\gamma}$ suggest that the only models in this class which would survive --- should the Generalized 
Einstein-Aether model apply on the scales relevant to observations of gravitational waves --- are those with 
$c_{13}\equiv 0$ and hence $\Pi^{\rm T}_{\rm ds} \equiv 0$. If $1+w_{\rm de}=0$, it can be shown that if $c_{13}=0$ 
then the scalar sector will be observationally equivalent to a cosmological constant. This is because $c_{13}$ also 
sets $\Pi^{\rm S}_{\rm ds} = 0$. It is possible that if $1+w_{\rm ds} \not = 0$ then this equivalence will be broken 
and will lead to interesting observational consequences~\cite{Battye2017}. We also note the striking analogy 
for $\alpha_{\rm M}$ between $f(R)$ and $F({\cal K})$ models.

\medskip
\noindent (3) {\it Massive gravity theories}. Differently from the models above, these theories consider the graviton 
to be massive ($m_{\rm GW}\neq0$) and in general the mass could be a function of time (and space) but the scalar and 
vector sectors are unaffected by this choice \citep{deRham2014}. From the general equation describing the propagation 
of gravitational waves \citep{Gumrukcuoglu2012,Gumrukcuoglu2013}
\be\label{eqn:MG}
 \ddot{h} + (3+\alpha_{\rm M})H\dot{h} + \left[(1+\alpha_{\rm T}){\rm K}^2H^2+m^2_{\rm GW}\right]h = 0\,,
\ee
we can deduce a modification to the coefficients in the Horndeski model 
\be
 \delta C_{h} =-\frac{1}{2}\frac{m^2}{m_{\rm pl}^2}\frac{m^2_{\rm GW}}{H^2}\,.
\ee
The $\beta_i$ functions read: $\beta_{\rm M}=\alpha_{\rm M}$ and 
$\beta_{\rm T}=(1+\alpha_{\rm T}){\rm K}^2+{\rm M}_{\rm GW}^2$ where ${\rm M}_{\rm GW}=m_{\rm GW}/H$.

\medskip
\noindent (4) {\it Elastic dark energy models}. These models represent a generalisation of the perfect fluid approach 
to dark energy where the rigidity of the medium is taken into account. In their simplest formulation elastic dark 
energy models are analogous to massive gravity models albeit the mass term introduced is not linked to the graviton 
itself. It was shown that~\citep{Battye2007,Pearson2014}
\be
 8\pi G_{\rm N}P_{\rm ds}\Pi_{\rm de}^{\rm T} = \left(\frac{\mu}{m_{\rm pl}^2}+2aH\frac{\nu}{m_{\rm pl}^2}\right)
                                                (h_i-h)-a\frac{\nu}{m_{\rm pl}^2}\dot{h}\,,
\ee
where $\mu$, identified as the rigidity modulus, and $\nu$ as the viscosity, are parameters with dimensions $M^4$ and 
$M^3$ respectively. The previous expression reduces to what found in \cite{Battye2007} for $\nu=0$. The $C_i$ 
coefficients are
\be
 C_{\ddot{h}} = 0\,, \quad 
 C_{\dot{h}} = -\frac{a\nu}{m_{\rm pl}^2H}\,, \quad 
 C_{h} = -\frac{\mu+2aH\nu}{m_{\rm pl}^2H^2}\,.
\ee
The additional term $h_i$ takes into account the formation time of the elastic medium.

\section{Suppressing modified gravity effects in the scalar sector}\label{sect:suppression}

In this Appendix we will discuss the principle of applying the same approach to suppressing modified gravity effects as 
${\rm K}\rightarrow\infty$ in the gravitational waves sector to the scalar sector. First let us define some parameters 
commonly used to quantify deviations from Einstein gravity. We will use a metric of the form
\begin{equation}
 ds^2 = -(1+2\phi)c^2dt^2 + a(t)^2(1-2\psi)\delta_{ij}dx^idx^j\,.
\end{equation}
The two of the Einstein equations yield
\begin{eqnarray}
 && -\frac{2}{3}{\rm K}^2\psi = \sum_i\Omega_{i}\Delta_i = \mu_{\psi}(a,{\rm K})\Omega_{\rm m}\Delta_{\rm m}\;,\\
 && -\frac{1}{3}{\rm K}^2\left(\psi-\phi\right) = \sum_i\Omega_iw_i\Pi^{\rm S}_i\;,
\end{eqnarray}
where the summation is over $i={\rm m}$ and ${\rm ds}$ and the relative contributions to the critical density are 
$\Omega_i\equiv\Omega_i(a)$. In what follows we will assume that $\Pi_{\rm m}^{\rm S}\equiv 0$ which is the case for a 
perfect pressureless fluid. The Weyl potential $\Psi=\textstyle{\tfrac{1}{2}}(\phi+\psi)$ is the quantity which 
leads to a number of observational effects notably lensing and is often parameterized as
\be
 -\frac{2}{3}{\rm K}^2\Psi = \Sigma(a,{\rm K})\Omega_{\rm m}\Delta_{\rm m}\,.
\ee
The functions $\mu_{\Psi}$ and $\Sigma$ have been introduced to encode the effects of modifications to gravity. 
In principle other parameters can be used to describe this but they are all related to $\mu_{\Psi}$ and $\Sigma$; 
any two independent parameters are needed to fully describe the theory. One notable alternative often used is the 
gravitational slip
\be
 \eta(a,{\rm K}) = \frac{\psi}{\phi} = \left(1-2
                                       \frac{\sum_i\Omega_iw_i\Pi_i^{\rm S}}{\sum_i\Omega_i\Delta_i}
                                       \right)^{-1}\,,
\ee
while one can also define $\mu_{\phi}$ according to 
\be
-\frac{2}{3}{\rm K}^2\phi = \mu_{\phi}(a,{\rm K})\Omega_{\rm m}\Delta_{\rm m}\,,
\ee
where $\mu_{\psi}(a,{\rm K})=\eta(a,{\rm K})\mu_{\phi}(a,{\rm K})$ and $\Sigma(a,{\rm K}) = 
\textstyle{\frac{1}{2}}\mu_{\phi}(a,{\rm K})[1+\eta(a,{\rm K})]=
\textstyle{\frac{1}{2}}[\mu_{\phi}(a,{\rm K})+\mu_{\psi}(a,{\rm K})]$.
If the dark sector were to only comprise a cosmological constant then $\mu_{\Psi}=\mu_{\Phi}=\eta\equiv 1$ and 
$\Sigma=1$.

Using the equation of state approach in the scalar sector it is necessary to specify two functions and it has been 
argued that the natural ones to specify  are the entropy perturbation, $w_{\rm ds}\Gamma$, and the scalar anisotropic 
stress, $w_{\rm ds}\Pi_{\rm ds}^{\rm S}$ which are both gauge invariant. Since the perturbations are linear these 
functions must be linear functions of the other perturbation variables and can be written (using the Einstein and 
conservation equations to remove metric perturbations and time derivatives) as
\begin{eqnarray}
 w_{\rm ds}\Gamma_{\rm ds} = & & C_{\Gamma_{\rm ds}\Delta_{\rm ds}}\Delta_{\rm ds} + 
                                 C_{\Gamma_{\rm ds}\Theta_{\rm ds}}\Theta_{\rm ds} + 
                                 C_{\Gamma_{\rm ds}\Delta_{\rm m}}\Delta_{\rm m}\nonumber\\ 
                              && +C_{\Gamma_{\rm ds}\Theta_{\rm m}}\Theta_{\rm m} + 
                                 C_{\Gamma_{\rm ds}\Gamma_{\rm m}}\Gamma_{\rm m}\,,\cr\cr
 w_{\rm ds}\Pi_{\rm ds}^{\rm S} = && C_{\Pi^{\rm S}_{\rm ds}\Delta_{\rm ds}}\Delta_{\rm ds} + 
                                     C_{\Pi^{\rm S}_{\rm ds}\Theta_{\rm ds}}\Theta_{\rm ds} + 
                                     C_{\Pi^{\rm S}_{\rm ds}\Delta_{\rm m}}\Delta_{\rm m} \nonumber\\
                                 && +C_{\Pi^{\rm S}_{\rm ds}\Theta_{\rm m}}\Theta_{\rm m} + 
                                     C_{\Pi^{\rm S}_{\rm ds}\Pi_{\rm m}^{\rm S}}\Pi_{\rm m}^{\rm S}\,.
                                      \label{scalartheory}
\end{eqnarray}
Here, we are describing the system where $\Delta_{i}$ and $\Theta_{i}$ are density and velocity perturbations in the 
dark (ds) and matter (m) sectors using the same notation as in \cite{Battye2016a}. For completeness we have also 
included the entropy perturbation $\Gamma_{\rm m}$ and the anisotropic stress $\Pi_{\rm m}^{\rm S}$ for the matter 
component which are typically negligible in the regime relevant to observations of cosmic acceleration; in the 
subsequent discussions we will ignore these terms. The coefficients $(C_{ij})$ have been computed for 
$k$-essence~\cite{Cheung2008,Creminelli2009,Battye2012}, kinetic gravity braiding~\cite{Battye2013}, 
$f(R)$~\cite{Battye2016a}, Horndeski theories~\cite{Gleyzes2014}, generalised Einstein-Aether~\cite{Battye2017}, 
elastic dark energy~\cite{Battye2007,Pearson2014} and Lorentz-violating massive gravity models~\cite{Battye2013a}. In 
full generality they are free functions of the scale factor (and hence cosmic time) and scale via the wave number, 
usually entering as a $k^2$ term due to the presumed dependence on second order combinations of spatial derivatives.

In order to establish a relationship between $\Delta_{\rm m}$ and $\Delta_{\rm ds}$ we will now assume that the 
approach to understanding perturbations in the scalar sector which works in $f(R)$ models (see, for example, 
\cite{Battye2017a}) works in more general models. We would assume that this is a good approximation to a wide range of 
models, but not all cases. In particular, we will assume that one can ignore the contributions from $\Theta_{\rm m}$ 
and $\Theta_{\rm ds}$ in (\ref{scalartheory}) and construct a second order differential equation describing the 
evolution of $\Delta_{\rm ds}$ which is sourced by matter perturbations
\bea
 \ddot{\Delta}_{\rm ds} & + & \left(2-3w_{\rm de}-2C_{\Pi^{\rm S}_{\rm ds}\Delta_{\rm ds}}\right)H\dot{\Delta}_{\rm ds} 
 \nonumber\\
 & + & \frac{1}{3}(3w_{\rm ds}+2C_{\Pi^{\rm S}_{\rm ds}\Delta_{\rm ds}}+3C_{\Gamma_{\rm ds}\Delta_{\rm ds}})
 H^2{\rm K}^2\Delta_{\rm ds} = \nonumber\\
 & -& \frac{1}{3}
 (2C_{\Pi^{\rm S}_{\rm ds}\Delta_{\rm m}}+3C_{\Gamma_{\rm ds}\Delta_{\rm m}})H^2{\rm K}^2\Delta_{\rm m}\;,
\eea
so that the relation between $\Delta_{\rm ds}$ and $\Delta_{\rm m}$ (the attractor solution) is
\be\label{eqn:attractor}
 \Delta_{\rm de} = 
 -\frac{2C_{\Pi^{\rm S}_{\rm ds}\Delta_{\rm m}}+3C_{\Gamma_{\rm ds}\Delta_{\rm m}}}
       {3w_{\rm ds}+2C_{\Pi^{\rm S}_{\rm ds}\Delta_{\rm ds}}+3C_{\Gamma_{\rm ds}\Delta_{\rm ds}}}\Delta_{\rm m}\;.
\ee
When this attractor solution applies, we can deduce that 
\bea
 \mu_{\psi} & = & 1-\frac{\Omega_{\rm ds}}{\Omega_{\rm m}}\kappa\,, \nonumber\\
 \Sigma & = & 1-\frac{\Omega_{\rm ds}}{\Omega_{\rm m}}\left[C_{\Pi^{\rm S}_{\rm ds}\Delta_{\rm m}} 
           +\kappa\left(1-C_{\Pi^{\rm S}_{\rm ds}\Delta_{\rm ds}}\right)\right]\,.
\eea
where we have defined
\be
 \kappa = \frac{2C_{\Pi^{\rm S}_{\rm ds}\Delta_{\rm m}}+3C_{\Gamma_{\rm ds}\Delta_{\rm m}}}
          {3w_{\rm ds}+2C_{\Pi^{\rm S}_{\rm ds}\Delta_{\rm ds}}+3C_{\Gamma_{\rm ds}\Delta_{\rm ds}}}\,.
\ee
Using these expressions, we see that one experiences general relativity on small scales if, in the limit 
${\rm K}\rightarrow\infty$, we have that $\mu_{\psi}\rightarrow 1$, which implies $\kappa\rightarrow 0$, and 
$\Sigma\rightarrow 1$, and this can be achieved if $\kappa=0$ and $C_{\Pi_{\rm ds}^{\rm S}}=0$. One specific way of 
enforcing this is by setting $C_{\Pi_{\rm ds}^{\rm S}\Delta_{\rm m}}=C_{\Gamma_{\rm ds}\Delta_{\rm m}} = 0$ in this 
limit, although there are other possible ways of achieving this. The suppression mechanism we are suggesting would 
require the zero in these conditions to be replaced by ${\cal O}({\rm K}^{-2})$ so that they are effectively zero for 
${\rm K}\sim {\rm K}_{\rm solar}\approx 4\times 10^{14}$. We will discuss the details of how this might be achieved in 
practice in future work. 

The attractor solution arises naturally when writing the equation for $\Delta_{\rm ds}$ with $\Delta_{\rm m}$ as source 
term. The expression in (\ref{eqn:attractor}) is valid provided that the attractor solution is attained for each 
Fourier mode before the dark energy component starts to dominate. Under the assumption that any modification of gravity 
is relevant only at late times, one would expect this will be true in the matter dominated era since we can in general 
assume that at very early times, i.e. in the radiation dominated era, perturbations in the dark sector are negligible. 
However, this need not to be the case when the dark energy component is not negligible at early times, such as in early 
dark energy models. In this case the attractor solution would have to be obtained during radiation dominated era and 
its validity would need to be checked carefully. If the field does not reach the attractor solution sufficiently fast 
to make exact initial conditions unimportant, then the full equations of motions need to be solved. For a deeper 
discussion on the issue we refer to \cite{Ballesteros2010} where this issue is discussed in detail for a perfect fluid.

It is interesting to calculate the expressions for $f(R)$ models since they exhibit some of the properties we are 
looking for, but not all. We will use approximations for the $C_{ij}$ coefficients presented in \cite{Battye2017a} 
which appear to give a good description of the full problem on all but the very largest scales when $f_R\ll 1$ (which 
one would expect to be the case),
\begin{fleqn}
 \bea
  && C_{\Pi_{\rm ds}^{\rm S}\Delta_{\rm m}} = 0 \,, \qquad \quad
     C_{\Pi^{\rm S}_{\rm ds}\Delta_{\rm ds}} = 1 \;, \nonumber\\
  && C_{\Gamma_{\rm ds}\Delta_{\rm m}} = \frac{1}{3}\frac{\Omega_{\rm m}}{\Omega_{\rm ds}}\,, \quad 
     C_{\Gamma_{\rm ds}\Delta_{\rm ds}} = \left(\frac{1}{3}-w_{\rm ds}+\frac{{\rm M}^2}{{\rm K}^2}\right)\,,
 \eea
\end{fleqn}
where ${\rm M}^2\equiv\dot{R}/(3H\dot{H}B)$ and $B=f_{RR}H\dot{R}/[\dot{H}(1+f_{R})]$ with $f_{R}=\tfrac{df}{dR}$ and 
$f_{RR}=\tfrac{d^2f}{dR^2}$. Using these expressions we can deduce that
\begin{fleqn}
 \bea
  && \kappa = \frac{1}{3}\frac{{\rm K}^2}{{\rm K}^2+{\rm M}^2}
                         \frac{\Omega_{\rm m}}{\Omega_{\rm ds}}\,, \quad 
     \mu_{\psi} = \frac{2{\rm K}^2+3{\rm M}^2}{3({\rm K}^2+{\rm M}^2)}\,, \nonumber\\
  && \mu_{\phi} = \frac{4{\rm K}^2+3{\rm M}^2}{3({\rm K}^2+{\rm M}^2)}\,, \quad 
     \eta =  \frac{2{\rm K}^2+3{\rm M}^2}{4{\rm K}^2+3{\rm M}^2}\,, \quad 
     \Sigma = 1\,.
 \eea
\end{fleqn}
In the small scale limit ${\rm K}\gg {\rm M}\gg 1$: $\mu_{\phi}\rightarrow4/3$ and this represents the well known
effective gravitational constant in $f(R)$ models leading to an increase of clustering on small scales; 
$\eta\rightarrow1/2$ and therefore the two Bardeen potentials differ from each other by a factor of two 
(another well known result \citep{DeFelice2010}). Finally we see at work the screening mechanism which reduces 
$\mu_{\psi}\rightarrow2/3$ and more importantly $\Sigma=1$ which is what would be expected for a model recovering 
general relativity on small scales, despite the fact that the $f(R)$ does not. This model achieves $\Sigma=1$ by having 
$C_{\Pi_{\rm ds}^{\rm S}\Delta_{\rm m}}=0$ and $C_{\Pi_{\rm ds}^{\rm S}\Delta_{\rm ds}}=1$ and does not have $\kappa=0$ 
and hence $\mu_{\psi}\ne 1$ and indeed it is believed that the $f(R)$ model can be compatible with the solar system 
scales by the nonlinear chameleon mechanism~\cite{Brax2008}.

\bibliographystyle{apsrev4-1}
\bibliography{gravwaves.bbl}

\label{lastpage}

\end{document}

%% file: defs.tex
\def\be{\begin{equation}}
\def\ee{\end{equation}}
\def\bea{\begin{eqnarray}}
\def\eea{\end{eqnarray}}
\def\bse{\begin{subequations}}
\def\ese{\end{subequations}}

\let\oldsqrt\sqrt
\def\sqrt{\mathpalette\DHLhksqrt}
\def\DHLhksqrt#1#2{%
\setbox0=\hbox{$#1\oldsqrt{#2\,}$}\dimen0=\ht0
\advance\dimen0-0.2\ht0
\setbox2=\hbox{\vrule height\ht0 depth -\dimen0}%
{\box0\lower0.4pt\box2}}

%% file: gravwaves.bbl
\begin{thebibliography}{75}%
\makeatletter
\providecommand \@ifxundefined [1]{%
 \@ifx{#1\undefined}
}%
\providecommand \@ifnum [1]{%
 \ifnum #1\expandafter \@firstoftwo
 \else \expandafter \@secondoftwo
 \fi
}%
\providecommand \@ifx [1]{%
 \ifx #1\expandafter \@firstoftwo
 \else \expandafter \@secondoftwo
 \fi
}%
\providecommand \natexlab [1]{#1}%
\providecommand \enquote  [1]{``#1''}%
\providecommand \bibnamefont  [1]{#1}%
\providecommand \bibfnamefont [1]{#1}%
\providecommand \citenamefont [1]{#1}%
\providecommand \href@noop [0]{\@secondoftwo}%
\providecommand \href [0]{\begingroup \@sanitize@url \@href}%
\providecommand \@href[1]{\@@startlink{#1}\@@href}%
\providecommand \@@href[1]{\endgroup#1\@@endlink}%
\providecommand \@sanitize@url [0]{\catcode `\\12\catcode `\$12\catcode
  `\&12\catcode `\#12\catcode `\^12\catcode `\_12\catcode `\%12\relax}%
\providecommand \@@startlink[1]{}%
\providecommand \@@endlink[0]{}%
\providecommand \url  [0]{\begingroup\@sanitize@url \@url }%
\providecommand \@url [1]{\endgroup\@href {#1}{\urlprefix }}%
\providecommand \urlprefix  [0]{URL }%
\providecommand \Eprint [0]{\href }%
\providecommand \doibase [0]{http://dx.doi.org/}%
\providecommand \selectlanguage [0]{\@gobble}%
\providecommand \bibinfo  [0]{\@secondoftwo}%
\providecommand \bibfield  [0]{\@secondoftwo}%
\providecommand \translation [1]{[#1]}%
\providecommand \BibitemOpen [0]{}%
\providecommand \bibitemStop [0]{}%
\providecommand \bibitemNoStop [0]{.\EOS\space}%
\providecommand \EOS [0]{\spacefactor3000\relax}%
\providecommand \BibitemShut  [1]{\csname bibitem#1\endcsname}%
\let\auto@bib@innerbib\@empty
\bibitem [{\citenamefont {{Abbott}}\ \emph
  {et~al.}(2017{\natexlab{a}})\citenamefont {{Abbott}}, \citenamefont
  {{Abbott}}, \citenamefont {{Abbott}}, \citenamefont {{Acernese}},
  \citenamefont {{Ackley}}, \citenamefont {{Adams}}, \citenamefont {{Adams}},
  \citenamefont {{Addesso}}, \citenamefont {{Adhikari}},\ and\ \citenamefont
  {{et~al.}}}]{LigoVirgo2017}%
  \BibitemOpen
  \bibfield  {author} {\bibinfo {author} {\bibfnamefont {B.~P.}\ \bibnamefont
  {{Abbott}}}, \bibinfo {author} {\bibfnamefont {R.}~\bibnamefont {{Abbott}}},
  \bibinfo {author} {\bibfnamefont {T.~D.}\ \bibnamefont {{Abbott}}}, \bibinfo
  {author} {\bibfnamefont {F.}~\bibnamefont {{Acernese}}}, \bibinfo {author}
  {\bibfnamefont {K.}~\bibnamefont {{Ackley}}}, \bibinfo {author}
  {\bibfnamefont {C.}~\bibnamefont {{Adams}}}, \bibinfo {author} {\bibfnamefont
  {T.}~\bibnamefont {{Adams}}}, \bibinfo {author} {\bibfnamefont
  {P.}~\bibnamefont {{Addesso}}}, \bibinfo {author} {\bibfnamefont {R.~X.}\
  \bibnamefont {{Adhikari}}}, \ and\ \bibinfo {author} {\bibnamefont
  {{et~al.}}},\ }\href {\doibase 10.1103/PhysRevLett.119.161101} {\bibfield
  {journal} {\bibinfo  {journal} {Physical Review Letters}\ }\textbf {\bibinfo
  {volume} {119}},\ \bibinfo {eid} {161101} (\bibinfo {year}
  {2017}{\natexlab{a}})},\ \Eprint {http://arxiv.org/abs/1710.05832}
  {arXiv:1710.05832 [gr-qc]} \BibitemShut {NoStop}%
\bibitem [{\citenamefont {{Abbott}}\ \emph
  {et~al.}(2017{\natexlab{b}})\citenamefont {{Abbott}}, \citenamefont
  {{Abbott}}, \citenamefont {{Abbott}}, \citenamefont {{Acernese}},
  \citenamefont {{Ackley}}, \citenamefont {{Adams}}, \citenamefont {{Adams}},
  \citenamefont {{Addesso}}, \citenamefont {{Adhikari}},\ and\ \citenamefont
  {{et~al.}}}]{LigoVirgoIntegral2017}%
  \BibitemOpen
  \bibfield  {author} {\bibinfo {author} {\bibfnamefont {B.~P.}\ \bibnamefont
  {{Abbott}}}, \bibinfo {author} {\bibfnamefont {R.}~\bibnamefont {{Abbott}}},
  \bibinfo {author} {\bibfnamefont {T.~D.}\ \bibnamefont {{Abbott}}}, \bibinfo
  {author} {\bibfnamefont {F.}~\bibnamefont {{Acernese}}}, \bibinfo {author}
  {\bibfnamefont {K.}~\bibnamefont {{Ackley}}}, \bibinfo {author}
  {\bibfnamefont {C.}~\bibnamefont {{Adams}}}, \bibinfo {author} {\bibfnamefont
  {T.}~\bibnamefont {{Adams}}}, \bibinfo {author} {\bibfnamefont
  {P.}~\bibnamefont {{Addesso}}}, \bibinfo {author} {\bibfnamefont {R.~X.}\
  \bibnamefont {{Adhikari}}}, \ and\ \bibinfo {author} {\bibnamefont
  {{et~al.}}},\ }\href {\doibase 10.3847/2041-8213/aa920c} {\bibfield
  {journal} {\bibinfo  {journal} {\apjl}\ }\textbf {\bibinfo {volume} {848}},\
  \bibinfo {eid} {L13} (\bibinfo {year} {2017}{\natexlab{b}})},\ \Eprint
  {http://arxiv.org/abs/1710.05834} {arXiv:1710.05834 [astro-ph.HE]}
  \BibitemShut {NoStop}%
\bibitem [{\citenamefont {{Abbott}}\ \emph
  {et~al.}(2017{\natexlab{c}})\citenamefont {{Abbott}}, \citenamefont
  {{Abbott}}, \citenamefont {{Abbott}}, \citenamefont {{Acernese}},
  \citenamefont {{Ackley}}, \citenamefont {{Adams}}, \citenamefont {{Adams}},
  \citenamefont {{Addesso}}, \citenamefont {{Adhikari}},\ and\ \citenamefont
  {{et~al.}}}]{LigoVirgo2017a}%
  \BibitemOpen
  \bibfield  {author} {\bibinfo {author} {\bibfnamefont {B.~P.}\ \bibnamefont
  {{Abbott}}}, \bibinfo {author} {\bibfnamefont {R.}~\bibnamefont {{Abbott}}},
  \bibinfo {author} {\bibfnamefont {T.~D.}\ \bibnamefont {{Abbott}}}, \bibinfo
  {author} {\bibfnamefont {F.}~\bibnamefont {{Acernese}}}, \bibinfo {author}
  {\bibfnamefont {K.}~\bibnamefont {{Ackley}}}, \bibinfo {author}
  {\bibfnamefont {C.}~\bibnamefont {{Adams}}}, \bibinfo {author} {\bibfnamefont
  {T.}~\bibnamefont {{Adams}}}, \bibinfo {author} {\bibfnamefont
  {P.}~\bibnamefont {{Addesso}}}, \bibinfo {author} {\bibfnamefont {R.~X.}\
  \bibnamefont {{Adhikari}}}, \ and\ \bibinfo {author} {\bibnamefont
  {{et~al.}}},\ }\href {\doibase 10.1038/nature24471} {\bibfield  {journal}
  {\bibinfo  {journal} {\nat}\ }\textbf {\bibinfo {volume} {551}},\ \bibinfo
  {pages} {85} (\bibinfo {year} {2017}{\natexlab{c}})},\ \Eprint
  {http://arxiv.org/abs/1710.05835} {arXiv:1710.05835} \BibitemShut {NoStop}%
\bibitem [{\citenamefont {{Caves}}(1980)}]{Caves1980}%
  \BibitemOpen
  \bibfield  {author} {\bibinfo {author} {\bibfnamefont {C.~M.}\ \bibnamefont
  {{Caves}}},\ }\href {\doibase 10.1016/0003-4916(80)90117-7} {\bibfield
  {journal} {\bibinfo  {journal} {Annals of Physics}\ }\textbf {\bibinfo
  {volume} {125}},\ \bibinfo {pages} {35} (\bibinfo {year} {1980})}\BibitemShut
  {NoStop}%
\bibitem [{\citenamefont {{Moore}}\ and\ \citenamefont
  {{Nelson}}(2001)}]{Moore2001}%
  \BibitemOpen
  \bibfield  {author} {\bibinfo {author} {\bibfnamefont {G.~D.}\ \bibnamefont
  {{Moore}}}\ and\ \bibinfo {author} {\bibfnamefont {A.~E.}\ \bibnamefont
  {{Nelson}}},\ }\href {\doibase 10.1088/1126-6708/2001/09/023} {\bibfield
  {journal} {\bibinfo  {journal} {Journal of High Energy Physics}\ }\textbf
  {\bibinfo {volume} {9}},\ \bibinfo {eid} {023} (\bibinfo {year} {2001})},\
  \Eprint {http://arxiv.org/abs/hep-ph/0106220} {hep-ph/0106220} \BibitemShut
  {NoStop}%
\bibitem [{\citenamefont {{Elliott}}\ \emph {et~al.}(2005)\citenamefont
  {{Elliott}}, \citenamefont {{Moore}},\ and\ \citenamefont
  {{Stoica}}}]{Elliott2005}%
  \BibitemOpen
  \bibfield  {author} {\bibinfo {author} {\bibfnamefont {J.~W.}\ \bibnamefont
  {{Elliott}}}, \bibinfo {author} {\bibfnamefont {G.~D.}\ \bibnamefont
  {{Moore}}}, \ and\ \bibinfo {author} {\bibfnamefont {H.}~\bibnamefont
  {{Stoica}}},\ }\href {\doibase 10.1088/1126-6708/2005/08/066} {\bibfield
  {journal} {\bibinfo  {journal} {Journal of High Energy Physics}\ }\textbf
  {\bibinfo {volume} {8}},\ \bibinfo {eid} {066} (\bibinfo {year} {2005})},\
  \Eprint {http://arxiv.org/abs/hep-ph/0505211} {hep-ph/0505211} \BibitemShut
  {NoStop}%
\bibitem [{\citenamefont {{Kimura}}\ and\ \citenamefont
  {{Yamamoto}}(2012)}]{Kimura2012}%
  \BibitemOpen
  \bibfield  {author} {\bibinfo {author} {\bibfnamefont {R.}~\bibnamefont
  {{Kimura}}}\ and\ \bibinfo {author} {\bibfnamefont {K.}~\bibnamefont
  {{Yamamoto}}},\ }\href {\doibase 10.1088/1475-7516/2012/07/050} {\bibfield
  {journal} {\bibinfo  {journal} {\jcap}\ }\textbf {\bibinfo {volume} {7}},\
  \bibinfo {eid} {050} (\bibinfo {year} {2012})},\ \Eprint
  {http://arxiv.org/abs/1112.4284} {arXiv:1112.4284 [astro-ph.CO]} \BibitemShut
  {NoStop}%
\bibitem [{\citenamefont {{Creminelli}}\ and\ \citenamefont
  {{Vernizzi}}(2017)}]{Creminelli2017}%
  \BibitemOpen
  \bibfield  {author} {\bibinfo {author} {\bibfnamefont {P.}~\bibnamefont
  {{Creminelli}}}\ and\ \bibinfo {author} {\bibfnamefont {F.}~\bibnamefont
  {{Vernizzi}}},\ }\href {\doibase 10.1103/PhysRevLett.119.251302} {\bibfield
  {journal} {\bibinfo  {journal} {Physical Review Letters}\ }\textbf {\bibinfo
  {volume} {119}},\ \bibinfo {eid} {251302} (\bibinfo {year} {2017})},\ \Eprint
  {http://arxiv.org/abs/1710.05877} {arXiv:1710.05877} \BibitemShut {NoStop}%
\bibitem [{\citenamefont {{Sakstein}}\ and\ \citenamefont
  {{Jain}}(2017)}]{Sakstein2017}%
  \BibitemOpen
  \bibfield  {author} {\bibinfo {author} {\bibfnamefont {J.}~\bibnamefont
  {{Sakstein}}}\ and\ \bibinfo {author} {\bibfnamefont {B.}~\bibnamefont
  {{Jain}}},\ }\href {\doibase 10.1103/PhysRevLett.119.251303} {\bibfield
  {journal} {\bibinfo  {journal} {Physical Review Letters}\ }\textbf {\bibinfo
  {volume} {119}},\ \bibinfo {eid} {251303} (\bibinfo {year} {2017})},\ \Eprint
  {http://arxiv.org/abs/1710.05893} {arXiv:1710.05893} \BibitemShut {NoStop}%
\bibitem [{\citenamefont {{Baker}}\ \emph {et~al.}(2017)\citenamefont
  {{Baker}}, \citenamefont {{Bellini}}, \citenamefont {{Ferreira}},
  \citenamefont {{Lagos}}, \citenamefont {{Noller}},\ and\ \citenamefont
  {{Sawicki}}}]{Baker2017}%
  \BibitemOpen
  \bibfield  {author} {\bibinfo {author} {\bibfnamefont {T.}~\bibnamefont
  {{Baker}}}, \bibinfo {author} {\bibfnamefont {E.}~\bibnamefont {{Bellini}}},
  \bibinfo {author} {\bibfnamefont {P.~G.}\ \bibnamefont {{Ferreira}}},
  \bibinfo {author} {\bibfnamefont {M.}~\bibnamefont {{Lagos}}}, \bibinfo
  {author} {\bibfnamefont {J.}~\bibnamefont {{Noller}}}, \ and\ \bibinfo
  {author} {\bibfnamefont {I.}~\bibnamefont {{Sawicki}}},\ }\href {\doibase
  10.1103/PhysRevLett.119.251301} {\bibfield  {journal} {\bibinfo  {journal}
  {Physical Review Letters}\ }\textbf {\bibinfo {volume} {119}},\ \bibinfo
  {eid} {251301} (\bibinfo {year} {2017})},\ \Eprint
  {http://arxiv.org/abs/1710.06394} {arXiv:1710.06394} \BibitemShut {NoStop}%
\bibitem [{\citenamefont {{Ezquiaga}}\ and\ \citenamefont
  {{Zumalac{\'a}rregui}}(2017)}]{Ezquiaga2017}%
  \BibitemOpen
  \bibfield  {author} {\bibinfo {author} {\bibfnamefont {J.~M.}\ \bibnamefont
  {{Ezquiaga}}}\ and\ \bibinfo {author} {\bibfnamefont {M.}~\bibnamefont
  {{Zumalac{\'a}rregui}}},\ }\href {\doibase 10.1103/PhysRevLett.119.251304}
  {\bibfield  {journal} {\bibinfo  {journal} {Physical Review Letters}\
  }\textbf {\bibinfo {volume} {119}},\ \bibinfo {eid} {251304} (\bibinfo {year}
  {2017})}\BibitemShut {NoStop}%
\bibitem [{\citenamefont {{Amendola}}\ \emph {et~al.}(2018)\citenamefont
  {{Amendola}}, \citenamefont {{Kunz}}, \citenamefont {{Saltas}},\ and\
  \citenamefont {{Sawicki}}}]{Amendola2018}%
  \BibitemOpen
  \bibfield  {author} {\bibinfo {author} {\bibfnamefont {L.}~\bibnamefont
  {{Amendola}}}, \bibinfo {author} {\bibfnamefont {M.}~\bibnamefont {{Kunz}}},
  \bibinfo {author} {\bibfnamefont {I.~D.}\ \bibnamefont {{Saltas}}}, \ and\
  \bibinfo {author} {\bibfnamefont {I.}~\bibnamefont {{Sawicki}}},\ }\href
  {\doibase 10.1103/PhysRevLett.120.131101} {\bibfield  {journal} {\bibinfo
  {journal} {Physical Review Letters}\ }\textbf {\bibinfo {volume} {120}},\
  \bibinfo {eid} {131101} (\bibinfo {year} {2018})},\ \Eprint
  {http://arxiv.org/abs/1711.04825} {arXiv:1711.04825} \BibitemShut {NoStop}%
\bibitem [{\citenamefont {{Crisostomi}}\ and\ \citenamefont
  {{Koyama}}(2018)}]{Crisostomi2018}%
  \BibitemOpen
  \bibfield  {author} {\bibinfo {author} {\bibfnamefont {M.}~\bibnamefont
  {{Crisostomi}}}\ and\ \bibinfo {author} {\bibfnamefont {K.}~\bibnamefont
  {{Koyama}}},\ }\href {\doibase 10.1103/PhysRevD.97.021301} {\bibfield
  {journal} {\bibinfo  {journal} {\prd}\ }\textbf {\bibinfo {volume} {97}},\
  \bibinfo {eid} {021301} (\bibinfo {year} {2018})},\ \Eprint
  {http://arxiv.org/abs/1711.06661} {arXiv:1711.06661} \BibitemShut {NoStop}%
\bibitem [{\citenamefont {{Nishizawa}}\ and\ \citenamefont
  {{Nakamura}}(2014)}]{Nishizawa2014}%
  \BibitemOpen
  \bibfield  {author} {\bibinfo {author} {\bibfnamefont {A.}~\bibnamefont
  {{Nishizawa}}}\ and\ \bibinfo {author} {\bibfnamefont {T.}~\bibnamefont
  {{Nakamura}}},\ }\href {\doibase 10.1103/PhysRevD.90.044048} {\bibfield
  {journal} {\bibinfo  {journal} {\prd}\ }\textbf {\bibinfo {volume} {90}},\
  \bibinfo {eid} {044048} (\bibinfo {year} {2014})},\ \Eprint
  {http://arxiv.org/abs/1406.5544} {arXiv:1406.5544 [gr-qc]} \BibitemShut
  {NoStop}%
\bibitem [{\citenamefont {{Lombriser}}\ and\ \citenamefont
  {{Taylor}}(2015)}]{Lombriser2015}%
  \BibitemOpen
  \bibfield  {author} {\bibinfo {author} {\bibfnamefont {L.}~\bibnamefont
  {{Lombriser}}}\ and\ \bibinfo {author} {\bibfnamefont {A.}~\bibnamefont
  {{Taylor}}},\ }\href {\doibase 10.1103/PhysRevLett.114.031101} {\bibfield
  {journal} {\bibinfo  {journal} {Physical Review Letters}\ }\textbf {\bibinfo
  {volume} {114}},\ \bibinfo {eid} {031101} (\bibinfo {year} {2015})},\ \Eprint
  {http://arxiv.org/abs/1405.2896} {arXiv:1405.2896} \BibitemShut {NoStop}%
\bibitem [{\citenamefont {{Lombriser}}\ and\ \citenamefont
  {{Taylor}}(2016)}]{Lombriser2016a}%
  \BibitemOpen
  \bibfield  {author} {\bibinfo {author} {\bibfnamefont {L.}~\bibnamefont
  {{Lombriser}}}\ and\ \bibinfo {author} {\bibfnamefont {A.}~\bibnamefont
  {{Taylor}}},\ }\href {\doibase 10.1088/1475-7516/2016/03/031} {\bibfield
  {journal} {\bibinfo  {journal} {\jcap}\ }\textbf {\bibinfo {volume} {3}},\
  \bibinfo {eid} {031} (\bibinfo {year} {2016})},\ \Eprint
  {http://arxiv.org/abs/1509.08458} {arXiv:1509.08458} \BibitemShut {NoStop}%
\bibitem [{\citenamefont {{Nishizawa}}(2016)}]{Nishizawa2016}%
  \BibitemOpen
  \bibfield  {author} {\bibinfo {author} {\bibfnamefont {A.}~\bibnamefont
  {{Nishizawa}}},\ }\href {\doibase 10.1103/PhysRevD.93.124036} {\bibfield
  {journal} {\bibinfo  {journal} {\prd}\ }\textbf {\bibinfo {volume} {93}},\
  \bibinfo {eid} {124036} (\bibinfo {year} {2016})},\ \Eprint
  {http://arxiv.org/abs/1601.01072} {arXiv:1601.01072 [gr-qc]} \BibitemShut
  {NoStop}%
\bibitem [{\citenamefont {{Lombriser}}\ and\ \citenamefont
  {{Lima}}(2017)}]{Lombriser2017}%
  \BibitemOpen
  \bibfield  {author} {\bibinfo {author} {\bibfnamefont {L.}~\bibnamefont
  {{Lombriser}}}\ and\ \bibinfo {author} {\bibfnamefont {N.~A.}\ \bibnamefont
  {{Lima}}},\ }\href {\doibase 10.1016/j.physletb.2016.12.048} {\bibfield
  {journal} {\bibinfo  {journal} {Physics Letters B}\ }\textbf {\bibinfo
  {volume} {765}},\ \bibinfo {pages} {382} (\bibinfo {year} {2017})},\ \Eprint
  {http://arxiv.org/abs/1602.07670} {arXiv:1602.07670} \BibitemShut {NoStop}%
\bibitem [{\citenamefont {{Battye}}\ and\ \citenamefont
  {{Pearson}}(2013{\natexlab{a}})}]{Battye2013}%
  \BibitemOpen
  \bibfield  {author} {\bibinfo {author} {\bibfnamefont {R.~A.}\ \bibnamefont
  {{Battye}}}\ and\ \bibinfo {author} {\bibfnamefont {J.~A.}\ \bibnamefont
  {{Pearson}}},\ }\href {\doibase 10.1103/PhysRevD.88.061301} {\bibfield
  {journal} {\bibinfo  {journal} {\prd}\ }\textbf {\bibinfo {volume} {88}},\
  \bibinfo {eid} {061301} (\bibinfo {year} {2013}{\natexlab{a}})},\ \Eprint
  {http://arxiv.org/abs/1306.1175} {arXiv:1306.1175} \BibitemShut {NoStop}%
\bibitem [{\citenamefont {{Battye}}\ and\ \citenamefont
  {{Pearson}}(2014)}]{Battye2014}%
  \BibitemOpen
  \bibfield  {author} {\bibinfo {author} {\bibfnamefont {R.~A.}\ \bibnamefont
  {{Battye}}}\ and\ \bibinfo {author} {\bibfnamefont {J.~A.}\ \bibnamefont
  {{Pearson}}},\ }\href {\doibase 10.1088/1475-7516/2014/03/051} {\bibfield
  {journal} {\bibinfo  {journal} {\jcap}\ }\textbf {\bibinfo {volume} {3}},\
  \bibinfo {eid} {051} (\bibinfo {year} {2014})},\ \Eprint
  {http://arxiv.org/abs/1311.6737} {arXiv:1311.6737} \BibitemShut {NoStop}%
\bibitem [{\citenamefont {{Battye}}\ \emph {et~al.}(2015)\citenamefont
  {{Battye}}, \citenamefont {{Moss}},\ and\ \citenamefont
  {{Pearson}}}]{Battye2015}%
  \BibitemOpen
  \bibfield  {author} {\bibinfo {author} {\bibfnamefont {R.~A.}\ \bibnamefont
  {{Battye}}}, \bibinfo {author} {\bibfnamefont {A.}~\bibnamefont {{Moss}}}, \
  and\ \bibinfo {author} {\bibfnamefont {J.~A.}\ \bibnamefont {{Pearson}}},\
  }\href {\doibase 10.1088/1475-7516/2015/04/048} {\bibfield  {journal}
  {\bibinfo  {journal} {\jcap}\ }\textbf {\bibinfo {volume} {4}},\ \bibinfo
  {eid} {048} (\bibinfo {year} {2015})},\ \Eprint
  {http://arxiv.org/abs/1409.4650} {arXiv:1409.4650} \BibitemShut {NoStop}%
\bibitem [{\citenamefont {{Soergel}}\ \emph {et~al.}(2015)\citenamefont
  {{Soergel}}, \citenamefont {{Giannantonio}}, \citenamefont {{Weller}},\ and\
  \citenamefont {{Battye}}}]{Soergel2015}%
  \BibitemOpen
  \bibfield  {author} {\bibinfo {author} {\bibfnamefont {B.}~\bibnamefont
  {{Soergel}}}, \bibinfo {author} {\bibfnamefont {T.}~\bibnamefont
  {{Giannantonio}}}, \bibinfo {author} {\bibfnamefont {J.}~\bibnamefont
  {{Weller}}}, \ and\ \bibinfo {author} {\bibfnamefont {R.~A.}\ \bibnamefont
  {{Battye}}},\ }\href {\doibase 10.1088/1475-7516/2015/02/037} {\bibfield
  {journal} {\bibinfo  {journal} {\jcap}\ }\textbf {\bibinfo {volume} {2}},\
  \bibinfo {eid} {037} (\bibinfo {year} {2015})},\ \Eprint
  {http://arxiv.org/abs/1409.4540} {arXiv:1409.4540} \BibitemShut {NoStop}%
\bibitem [{\citenamefont {{Nishizawa}}(2017)}]{Nishizawa2017}%
  \BibitemOpen
  \bibfield  {author} {\bibinfo {author} {\bibfnamefont {A.}~\bibnamefont
  {{Nishizawa}}},\ }\href@noop {} {\bibfield  {journal} {\bibinfo  {journal}
  {ArXiv e-prints}\ } (\bibinfo {year} {2017})},\ \Eprint
  {http://arxiv.org/abs/1710.04825} {arXiv:1710.04825 [gr-qc]} \BibitemShut
  {NoStop}%
\bibitem [{\citenamefont {{Arai}}\ and\ \citenamefont
  {{Nishizawa}}(2017)}]{Arai2017}%
  \BibitemOpen
  \bibfield  {author} {\bibinfo {author} {\bibfnamefont {S.}~\bibnamefont
  {{Arai}}}\ and\ \bibinfo {author} {\bibfnamefont {A.}~\bibnamefont
  {{Nishizawa}}},\ }\href@noop {} {\bibfield  {journal} {\bibinfo  {journal}
  {ArXiv e-prints}\ } (\bibinfo {year} {2017})},\ \Eprint
  {http://arxiv.org/abs/1711.03776} {arXiv:1711.03776 [gr-qc]} \BibitemShut
  {NoStop}%
\bibitem [{\citenamefont {{Balek}}\ and\ \citenamefont
  {{Pol{\'a}k}}(2009)}]{Balek2009}%
  \BibitemOpen
  \bibfield  {author} {\bibinfo {author} {\bibfnamefont {V.}~\bibnamefont
  {{Balek}}}\ and\ \bibinfo {author} {\bibfnamefont {V.}~\bibnamefont
  {{Pol{\'a}k}}},\ }\href {\doibase 10.1007/s10714-008-0681-x} {\bibfield
  {journal} {\bibinfo  {journal} {General Relativity and Gravitation}\ }\textbf
  {\bibinfo {volume} {41}},\ \bibinfo {pages} {505} (\bibinfo {year} {2009})},\
  \Eprint {http://arxiv.org/abs/0707.1513} {arXiv:0707.1513 [gr-qc]}
  \BibitemShut {NoStop}%
\bibitem [{\citenamefont {{Amendola}}\ \emph {et~al.}(2017)\citenamefont
  {{Amendola}}, \citenamefont {{Sawicki}}, \citenamefont {{Kunz}},\ and\
  \citenamefont {{Saltas}}}]{Amendola2017b}%
  \BibitemOpen
  \bibfield  {author} {\bibinfo {author} {\bibfnamefont {L.}~\bibnamefont
  {{Amendola}}}, \bibinfo {author} {\bibfnamefont {I.}~\bibnamefont
  {{Sawicki}}}, \bibinfo {author} {\bibfnamefont {M.}~\bibnamefont {{Kunz}}}, \
  and\ \bibinfo {author} {\bibfnamefont {I.~D.}\ \bibnamefont {{Saltas}}},\
  }\href@noop {} {\bibfield  {journal} {\bibinfo  {journal} {ArXiv e-prints}\ }
  (\bibinfo {year} {2017})},\ \Eprint {http://arxiv.org/abs/1712.08623}
  {arXiv:1712.08623} \BibitemShut {NoStop}%
\bibitem [{\citenamefont {{Bertotti}}\ \emph {et~al.}(2003)\citenamefont
  {{Bertotti}}, \citenamefont {{Iess}},\ and\ \citenamefont
  {{Tortora}}}]{Bertotti2003}%
  \BibitemOpen
  \bibfield  {author} {\bibinfo {author} {\bibfnamefont {B.}~\bibnamefont
  {{Bertotti}}}, \bibinfo {author} {\bibfnamefont {L.}~\bibnamefont {{Iess}}},
  \ and\ \bibinfo {author} {\bibfnamefont {P.}~\bibnamefont {{Tortora}}},\
  }\href {\doibase 10.1038/nature01997} {\bibfield  {journal} {\bibinfo
  {journal} {\nat}\ }\textbf {\bibinfo {volume} {425}},\ \bibinfo {pages} {374}
  (\bibinfo {year} {2003})}\BibitemShut {NoStop}%
\bibitem [{\citenamefont {{Will}}(2006)}]{Will2006}%
  \BibitemOpen
  \bibfield  {author} {\bibinfo {author} {\bibfnamefont {C.~M.}\ \bibnamefont
  {{Will}}},\ }\href {\doibase 10.12942/lrr-2006-3} {\bibfield  {journal}
  {\bibinfo  {journal} {Living Reviews in Relativity}\ }\textbf {\bibinfo
  {volume} {9}},\ \bibinfo {eid} {3} (\bibinfo {year} {2006})},\ \Eprint
  {http://arxiv.org/abs/gr-qc/0510072} {gr-qc/0510072} \BibitemShut {NoStop}%
\bibitem [{\citenamefont {{Abbott}}\ \emph {et~al.}(2016)\citenamefont
  {{Abbott}}, \citenamefont {{Abbott}}, \citenamefont {{Abbott}}, \citenamefont
  {{Abernathy}}, \citenamefont {{Acernese}}, \citenamefont {{Ackley}},
  \citenamefont {{Adams}}, \citenamefont {{Adams}}, \citenamefont {{Addesso}},\
  and\ \citenamefont {{et~al.}}}]{LigoVirgo2016}%
  \BibitemOpen
  \bibfield  {author} {\bibinfo {author} {\bibfnamefont {B.~P.}\ \bibnamefont
  {{Abbott}}}, \bibinfo {author} {\bibfnamefont {R.}~\bibnamefont {{Abbott}}},
  \bibinfo {author} {\bibfnamefont {T.~D.}\ \bibnamefont {{Abbott}}}, \bibinfo
  {author} {\bibfnamefont {M.~R.}\ \bibnamefont {{Abernathy}}}, \bibinfo
  {author} {\bibfnamefont {F.}~\bibnamefont {{Acernese}}}, \bibinfo {author}
  {\bibfnamefont {K.}~\bibnamefont {{Ackley}}}, \bibinfo {author}
  {\bibfnamefont {C.}~\bibnamefont {{Adams}}}, \bibinfo {author} {\bibfnamefont
  {T.}~\bibnamefont {{Adams}}}, \bibinfo {author} {\bibfnamefont
  {P.}~\bibnamefont {{Addesso}}}, \ and\ \bibinfo {author} {\bibnamefont
  {{et~al.}}},\ }\href {\doibase 10.1103/PhysRevLett.116.221101} {\bibfield
  {journal} {\bibinfo  {journal} {Physical Review Letters}\ }\textbf {\bibinfo
  {volume} {116}},\ \bibinfo {eid} {221101} (\bibinfo {year} {2016})},\ \Eprint
  {http://arxiv.org/abs/1602.03841} {arXiv:1602.03841 [gr-qc]} \BibitemShut
  {NoStop}%
\bibitem [{\citenamefont {{de Rham}}\ \emph {et~al.}(2017)\citenamefont {{de
  Rham}}, \citenamefont {{Deskins}}, \citenamefont {{Tolley}},\ and\
  \citenamefont {{Zhou}}}]{deRham2017}%
  \BibitemOpen
  \bibfield  {author} {\bibinfo {author} {\bibfnamefont {C.}~\bibnamefont {{de
  Rham}}}, \bibinfo {author} {\bibfnamefont {J.~T.}\ \bibnamefont {{Deskins}}},
  \bibinfo {author} {\bibfnamefont {A.~J.}\ \bibnamefont {{Tolley}}}, \ and\
  \bibinfo {author} {\bibfnamefont {S.-Y.}\ \bibnamefont {{Zhou}}},\ }\href
  {\doibase 10.1103/RevModPhys.89.025004} {\bibfield  {journal} {\bibinfo
  {journal} {Reviews of Modern Physics}\ }\textbf {\bibinfo {volume} {89}},\
  \bibinfo {eid} {025004} (\bibinfo {year} {2017})},\ \Eprint
  {http://arxiv.org/abs/1606.08462} {arXiv:1606.08462} \BibitemShut {NoStop}%
\bibitem [{\citenamefont {{Choudhury}}\ \emph {et~al.}(2004)\citenamefont
  {{Choudhury}}, \citenamefont {{Joshi}}, \citenamefont {{Mahajan}},\ and\
  \citenamefont {{McKellar}}}]{Choudhury2004}%
  \BibitemOpen
  \bibfield  {author} {\bibinfo {author} {\bibfnamefont {S.~R.}\ \bibnamefont
  {{Choudhury}}}, \bibinfo {author} {\bibfnamefont {G.~C.}\ \bibnamefont
  {{Joshi}}}, \bibinfo {author} {\bibfnamefont {S.}~\bibnamefont {{Mahajan}}},
  \ and\ \bibinfo {author} {\bibfnamefont {B.~H.~J.}\ \bibnamefont
  {{McKellar}}},\ }\href {\doibase 10.1016/j.astropartphys.2004.04.001}
  {\bibfield  {journal} {\bibinfo  {journal} {Astroparticle Physics}\ }\textbf
  {\bibinfo {volume} {21}},\ \bibinfo {pages} {559} (\bibinfo {year} {2004})},\
  \Eprint {http://arxiv.org/abs/hep-ph/0204161} {hep-ph/0204161} \BibitemShut
  {NoStop}%
\bibitem [{\citenamefont {{Desai}}(2018)}]{Desai2018}%
  \BibitemOpen
  \bibfield  {author} {\bibinfo {author} {\bibfnamefont {S.}~\bibnamefont
  {{Desai}}},\ }\href {\doibase 10.1016/j.physletb.2018.01.052} {\bibfield
  {journal} {\bibinfo  {journal} {Physics Letters B}\ }\textbf {\bibinfo
  {volume} {778}},\ \bibinfo {pages} {325} (\bibinfo {year} {2018})},\ \Eprint
  {http://arxiv.org/abs/1708.06502} {arXiv:1708.06502} \BibitemShut {NoStop}%
\bibitem [{\citenamefont {{Rana}}\ \emph {et~al.}(2018)\citenamefont {{Rana}},
  \citenamefont {{Jain}}, \citenamefont {{Mahajan}},\ and\ \citenamefont
  {{Mukherjee}}}]{Rana2018}%
  \BibitemOpen
  \bibfield  {author} {\bibinfo {author} {\bibfnamefont {A.}~\bibnamefont
  {{Rana}}}, \bibinfo {author} {\bibfnamefont {D.}~\bibnamefont {{Jain}}},
  \bibinfo {author} {\bibfnamefont {S.}~\bibnamefont {{Mahajan}}}, \ and\
  \bibinfo {author} {\bibfnamefont {A.}~\bibnamefont {{Mukherjee}}},\ }\href
  {\doibase 10.1016/j.physletb.2018.03.076} {\bibfield  {journal} {\bibinfo
  {journal} {Physics Letters B}\ }\textbf {\bibinfo {volume} {781}},\ \bibinfo
  {pages} {220} (\bibinfo {year} {2018})},\ \Eprint
  {http://arxiv.org/abs/1801.03309} {arXiv:1801.03309} \BibitemShut {NoStop}%
\bibitem [{\citenamefont {{Bettoni}}\ \emph {et~al.}(2017)\citenamefont
  {{Bettoni}}, \citenamefont {{Ezquiaga}}, \citenamefont {{Hinterbichler}},\
  and\ \citenamefont {{Zumalac{\'a}rregui}}}]{Bettoni2017}%
  \BibitemOpen
  \bibfield  {author} {\bibinfo {author} {\bibfnamefont {D.}~\bibnamefont
  {{Bettoni}}}, \bibinfo {author} {\bibfnamefont {J.~M.}\ \bibnamefont
  {{Ezquiaga}}}, \bibinfo {author} {\bibfnamefont {K.}~\bibnamefont
  {{Hinterbichler}}}, \ and\ \bibinfo {author} {\bibfnamefont {M.}~\bibnamefont
  {{Zumalac{\'a}rregui}}},\ }\href {\doibase 10.1103/PhysRevD.95.084029}
  {\bibfield  {journal} {\bibinfo  {journal} {\prd}\ }\textbf {\bibinfo
  {volume} {95}},\ \bibinfo {eid} {084029} (\bibinfo {year} {2017})},\ \Eprint
  {http://arxiv.org/abs/1608.01982} {arXiv:1608.01982 [gr-qc]} \BibitemShut
  {NoStop}%
\bibitem [{\citenamefont {{Maggiore}}\ and\ \citenamefont
  {{Mancarella}}(2014)}]{Maggiore2014}%
  \BibitemOpen
  \bibfield  {author} {\bibinfo {author} {\bibfnamefont {M.}~\bibnamefont
  {{Maggiore}}}\ and\ \bibinfo {author} {\bibfnamefont {M.}~\bibnamefont
  {{Mancarella}}},\ }\href {\doibase 10.1103/PhysRevD.90.023005} {\bibfield
  {journal} {\bibinfo  {journal} {\prd}\ }\textbf {\bibinfo {volume} {90}},\
  \bibinfo {eid} {023005} (\bibinfo {year} {2014})},\ \Eprint
  {http://arxiv.org/abs/1402.0448} {arXiv:1402.0448 [hep-th]} \BibitemShut
  {NoStop}%
\bibitem [{\citenamefont {{Belgacem}}\ \emph {et~al.}(2018)\citenamefont
  {{Belgacem}}, \citenamefont {{Dirian}}, \citenamefont {{Foffa}},\ and\
  \citenamefont {{Maggiore}}}]{Belgacem2018}%
  \BibitemOpen
  \bibfield  {author} {\bibinfo {author} {\bibfnamefont {E.}~\bibnamefont
  {{Belgacem}}}, \bibinfo {author} {\bibfnamefont {Y.}~\bibnamefont
  {{Dirian}}}, \bibinfo {author} {\bibfnamefont {S.}~\bibnamefont {{Foffa}}}, \
  and\ \bibinfo {author} {\bibfnamefont {M.}~\bibnamefont {{Maggiore}}},\
  }\href {\doibase 10.1088/1475-7516/2018/03/002} {\bibfield  {journal}
  {\bibinfo  {journal} {\jcap}\ }\textbf {\bibinfo {volume} {3}},\ \bibinfo
  {eid} {002} (\bibinfo {year} {2018})},\ \Eprint
  {http://arxiv.org/abs/1712.07066} {arXiv:1712.07066 [hep-th]} \BibitemShut
  {NoStop}%
\bibitem [{\citenamefont {{Chicone}}\ and\ \citenamefont
  {{Mashhoon}}(2013)}]{Chicone2013}%
  \BibitemOpen
  \bibfield  {author} {\bibinfo {author} {\bibfnamefont {C.}~\bibnamefont
  {{Chicone}}}\ and\ \bibinfo {author} {\bibfnamefont {B.}~\bibnamefont
  {{Mashhoon}}},\ }\href {\doibase 10.1103/PhysRevD.87.064015} {\bibfield
  {journal} {\bibinfo  {journal} {\prd}\ }\textbf {\bibinfo {volume} {87}},\
  \bibinfo {eid} {064015} (\bibinfo {year} {2013})},\ \Eprint
  {http://arxiv.org/abs/1210.3860} {arXiv:1210.3860 [gr-qc]} \BibitemShut
  {NoStop}%
\bibitem [{\citenamefont {{Mashhoon}}(2014)}]{Mashhoon2014}%
  \BibitemOpen
  \bibfield  {author} {\bibinfo {author} {\bibfnamefont {B.}~\bibnamefont
  {{Mashhoon}}},\ }\href {\doibase 10.1103/PhysRevD.90.124031} {\bibfield
  {journal} {\bibinfo  {journal} {\prd}\ }\textbf {\bibinfo {volume} {90}},\
  \bibinfo {eid} {124031} (\bibinfo {year} {2014})},\ \Eprint
  {http://arxiv.org/abs/1409.4472} {arXiv:1409.4472 [gr-qc]} \BibitemShut
  {NoStop}%
\bibitem [{\citenamefont {{Stelle}}(1978)}]{Stelle1978}%
  \BibitemOpen
  \bibfield  {author} {\bibinfo {author} {\bibfnamefont {K.~S.}\ \bibnamefont
  {{Stelle}}},\ }\href {\doibase 10.1007/BF00760427} {\bibfield  {journal}
  {\bibinfo  {journal} {General Relativity and Gravitation}\ }\textbf {\bibinfo
  {volume} {9}},\ \bibinfo {pages} {353} (\bibinfo {year} {1978})}\BibitemShut
  {NoStop}%
\bibitem [{\citenamefont {{Mirshekari}}\ \emph {et~al.}(2012)\citenamefont
  {{Mirshekari}}, \citenamefont {{Yunes}},\ and\ \citenamefont
  {{Will}}}]{Mirshekari2012}%
  \BibitemOpen
  \bibfield  {author} {\bibinfo {author} {\bibfnamefont {S.}~\bibnamefont
  {{Mirshekari}}}, \bibinfo {author} {\bibfnamefont {N.}~\bibnamefont
  {{Yunes}}}, \ and\ \bibinfo {author} {\bibfnamefont {C.~M.}\ \bibnamefont
  {{Will}}},\ }\href {\doibase 10.1103/PhysRevD.85.024041} {\bibfield
  {journal} {\bibinfo  {journal} {\prd}\ }\textbf {\bibinfo {volume} {85}},\
  \bibinfo {eid} {024041} (\bibinfo {year} {2012})},\ \Eprint
  {http://arxiv.org/abs/1110.2720} {arXiv:1110.2720 [gr-qc]} \BibitemShut
  {NoStop}%
\bibitem [{\citenamefont {{Yunes}}\ \emph {et~al.}(2016)\citenamefont
  {{Yunes}}, \citenamefont {{Yagi}},\ and\ \citenamefont
  {{Pretorius}}}]{Yunes2016}%
  \BibitemOpen
  \bibfield  {author} {\bibinfo {author} {\bibfnamefont {N.}~\bibnamefont
  {{Yunes}}}, \bibinfo {author} {\bibfnamefont {K.}~\bibnamefont {{Yagi}}}, \
  and\ \bibinfo {author} {\bibfnamefont {F.}~\bibnamefont {{Pretorius}}},\
  }\href {\doibase 10.1103/PhysRevD.94.084002} {\bibfield  {journal} {\bibinfo
  {journal} {\prd}\ }\textbf {\bibinfo {volume} {94}},\ \bibinfo {eid} {084002}
  (\bibinfo {year} {2016})},\ \Eprint {http://arxiv.org/abs/1603.08955}
  {arXiv:1603.08955 [gr-qc]} \BibitemShut {NoStop}%
\bibitem [{\citenamefont {{Samajdar}}\ and\ \citenamefont
  {{Arun}}(2017)}]{Samajdar2017}%
  \BibitemOpen
  \bibfield  {author} {\bibinfo {author} {\bibfnamefont {A.}~\bibnamefont
  {{Samajdar}}}\ and\ \bibinfo {author} {\bibfnamefont {K.~G.}\ \bibnamefont
  {{Arun}}},\ }\href {\doibase 10.1103/PhysRevD.96.104027} {\bibfield
  {journal} {\bibinfo  {journal} {\prd}\ }\textbf {\bibinfo {volume} {96}},\
  \bibinfo {eid} {104027} (\bibinfo {year} {2017})},\ \Eprint
  {http://arxiv.org/abs/1708.00671} {arXiv:1708.00671 [gr-qc]} \BibitemShut
  {NoStop}%
\bibitem [{\citenamefont {{Bellini}}\ and\ \citenamefont
  {{Sawicki}}(2014)}]{Bellini2014}%
  \BibitemOpen
  \bibfield  {author} {\bibinfo {author} {\bibfnamefont {E.}~\bibnamefont
  {{Bellini}}}\ and\ \bibinfo {author} {\bibfnamefont {I.}~\bibnamefont
  {{Sawicki}}},\ }\href {\doibase 10.1088/1475-7516/2014/07/050} {\bibfield
  {journal} {\bibinfo  {journal} {\jcap}\ }\textbf {\bibinfo {volume} {7}},\
  \bibinfo {eid} {050} (\bibinfo {year} {2014})},\ \Eprint
  {http://arxiv.org/abs/1404.3713} {arXiv:1404.3713} \BibitemShut {NoStop}%
\bibitem [{\citenamefont {{Gleyzes}}\ \emph {et~al.}(2013)\citenamefont
  {{Gleyzes}}, \citenamefont {{Langlois}}, \citenamefont {{Piazza}},\ and\
  \citenamefont {{Vernizzi}}}]{Gleyzes2013}%
  \BibitemOpen
  \bibfield  {author} {\bibinfo {author} {\bibfnamefont {J.}~\bibnamefont
  {{Gleyzes}}}, \bibinfo {author} {\bibfnamefont {D.}~\bibnamefont
  {{Langlois}}}, \bibinfo {author} {\bibfnamefont {F.}~\bibnamefont
  {{Piazza}}}, \ and\ \bibinfo {author} {\bibfnamefont {F.}~\bibnamefont
  {{Vernizzi}}},\ }\href {\doibase 10.1088/1475-7516/2013/08/025} {\bibfield
  {journal} {\bibinfo  {journal} {\jcap}\ }\textbf {\bibinfo {volume} {8}},\
  \bibinfo {eid} {025} (\bibinfo {year} {2013})},\ \Eprint
  {http://arxiv.org/abs/1304.4840} {arXiv:1304.4840 [hep-th]} \BibitemShut
  {NoStop}%
\bibitem [{\citenamefont {{Gleyzes}}\ \emph {et~al.}(2014)\citenamefont
  {{Gleyzes}}, \citenamefont {{Langlois}},\ and\ \citenamefont
  {{Vernizzi}}}]{Gleyzes2014}%
  \BibitemOpen
  \bibfield  {author} {\bibinfo {author} {\bibfnamefont {J.}~\bibnamefont
  {{Gleyzes}}}, \bibinfo {author} {\bibfnamefont {D.}~\bibnamefont
  {{Langlois}}}, \ and\ \bibinfo {author} {\bibfnamefont {F.}~\bibnamefont
  {{Vernizzi}}},\ }\href {\doibase 10.1142/S021827181443010X} {\bibfield
  {journal} {\bibinfo  {journal} {International Journal of Modern Physics D}\
  }\textbf {\bibinfo {volume} {23}},\ \bibinfo {eid} {1443010} (\bibinfo {year}
  {2014})},\ \Eprint {http://arxiv.org/abs/1411.3712} {arXiv:1411.3712
  [hep-th]} \BibitemShut {NoStop}%
\bibitem [{\citenamefont {{Tsujikawa}}(2015)}]{Tsujikawa2015}%
  \BibitemOpen
  \bibfield  {author} {\bibinfo {author} {\bibfnamefont {S.}~\bibnamefont
  {{Tsujikawa}}},\ }in\ \href {\doibase 10.1007/978-3-319-10070-8_4} {\emph
  {\bibinfo {booktitle} {Lecture Notes in Physics, Berlin Springer Verlag}}},\
  \bibinfo {series} {Lecture Notes in Physics, Berlin Springer Verlag}, Vol.\
  \bibinfo {volume} {892},\ \bibinfo {editor} {edited by\ \bibinfo {editor}
  {\bibfnamefont {E.}~\bibnamefont {{Papantonopoulos}}}}\ (\bibinfo {year}
  {2015})\ p.~\bibinfo {pages} {97},\ \Eprint {http://arxiv.org/abs/1404.2684}
  {arXiv:1404.2684 [gr-qc]} \BibitemShut {NoStop}%
\bibitem [{\citenamefont {{Ford}}(1987)}]{Ford1987}%
  \BibitemOpen
  \bibfield  {author} {\bibinfo {author} {\bibfnamefont {L.~H.}\ \bibnamefont
  {{Ford}}},\ }\href {\doibase 10.1103/PhysRevD.35.2339} {\bibfield  {journal}
  {\bibinfo  {journal} {\prd}\ }\textbf {\bibinfo {volume} {35}},\ \bibinfo
  {pages} {2339} (\bibinfo {year} {1987})}\BibitemShut {NoStop}%
\bibitem [{\citenamefont {{Peebles}}\ and\ \citenamefont
  {{Ratra}}(1988)}]{Peebles1988}%
  \BibitemOpen
  \bibfield  {author} {\bibinfo {author} {\bibfnamefont {P.~J.~E.}\
  \bibnamefont {{Peebles}}}\ and\ \bibinfo {author} {\bibfnamefont
  {B.}~\bibnamefont {{Ratra}}},\ }\href {\doibase 10.1086/185100} {\bibfield
  {journal} {\bibinfo  {journal} {\apjl}\ }\textbf {\bibinfo {volume} {325}},\
  \bibinfo {pages} {L17} (\bibinfo {year} {1988})}\BibitemShut {NoStop}%
\bibitem [{\citenamefont {{Ratra}}\ and\ \citenamefont
  {{Peebles}}(1988)}]{Ratra1988a}%
  \BibitemOpen
  \bibfield  {author} {\bibinfo {author} {\bibfnamefont {B.}~\bibnamefont
  {{Ratra}}}\ and\ \bibinfo {author} {\bibfnamefont {P.~J.~E.}\ \bibnamefont
  {{Peebles}}},\ }\href {\doibase 10.1103/PhysRevD.37.3406} {\bibfield
  {journal} {\bibinfo  {journal} {\prd}\ }\textbf {\bibinfo {volume} {37}},\
  \bibinfo {pages} {3406} (\bibinfo {year} {1988})}\BibitemShut {NoStop}%
\bibitem [{\citenamefont {{Wetterich}}(1988)}]{Wetterich1988}%
  \BibitemOpen
  \bibfield  {author} {\bibinfo {author} {\bibfnamefont {C.}~\bibnamefont
  {{Wetterich}}},\ }\href {\doibase 10.1016/0550-3213(88)90193-9} {\bibfield
  {journal} {\bibinfo  {journal} {Nuclear Physics B}\ }\textbf {\bibinfo
  {volume} {302}},\ \bibinfo {pages} {668} (\bibinfo {year}
  {1988})}\BibitemShut {NoStop}%
\bibitem [{\citenamefont {{Caldwell}}\ \emph {et~al.}(1998)\citenamefont
  {{Caldwell}}, \citenamefont {{Dave}},\ and\ \citenamefont
  {{Steinhardt}}}]{Caldwell1998}%
  \BibitemOpen
  \bibfield  {author} {\bibinfo {author} {\bibfnamefont {R.~R.}\ \bibnamefont
  {{Caldwell}}}, \bibinfo {author} {\bibfnamefont {R.}~\bibnamefont {{Dave}}},
  \ and\ \bibinfo {author} {\bibfnamefont {P.~J.}\ \bibnamefont
  {{Steinhardt}}},\ }\href {\doibase 10.1103/PhysRevLett.80.1582} {\bibfield
  {journal} {\bibinfo  {journal} {Physical Review Letters}\ }\textbf {\bibinfo
  {volume} {80}},\ \bibinfo {pages} {1582} (\bibinfo {year} {1998})},\ \Eprint
  {http://arxiv.org/abs/arXiv:astro-ph/9708069} {arXiv:astro-ph/9708069}
  \BibitemShut {NoStop}%
\bibitem [{\citenamefont {{Armend{\'a}riz-Pic{\'o}n}}\ \emph
  {et~al.}(1999)\citenamefont {{Armend{\'a}riz-Pic{\'o}n}}, \citenamefont
  {{Damour}},\ and\ \citenamefont {{Mukhanov}}}]{ArmendarizPicon1999}%
  \BibitemOpen
  \bibfield  {author} {\bibinfo {author} {\bibfnamefont {C.}~\bibnamefont
  {{Armend{\'a}riz-Pic{\'o}n}}}, \bibinfo {author} {\bibfnamefont
  {T.}~\bibnamefont {{Damour}}}, \ and\ \bibinfo {author} {\bibfnamefont
  {V.}~\bibnamefont {{Mukhanov}}},\ }\href {\doibase
  10.1016/S0370-2693(99)00603-6} {\bibfield  {journal} {\bibinfo  {journal}
  {Physics Letters B}\ }\textbf {\bibinfo {volume} {458}},\ \bibinfo {pages}
  {209} (\bibinfo {year} {1999})},\ \Eprint
  {http://arxiv.org/abs/hep-th/9904075} {hep-th/9904075} \BibitemShut {NoStop}%
\bibitem [{\citenamefont {{Chiba}}\ \emph {et~al.}(2000)\citenamefont
  {{Chiba}}, \citenamefont {{Okabe}},\ and\ \citenamefont
  {{Yamaguchi}}}]{Chiba2000}%
  \BibitemOpen
  \bibfield  {author} {\bibinfo {author} {\bibfnamefont {T.}~\bibnamefont
  {{Chiba}}}, \bibinfo {author} {\bibfnamefont {T.}~\bibnamefont {{Okabe}}}, \
  and\ \bibinfo {author} {\bibfnamefont {M.}~\bibnamefont {{Yamaguchi}}},\
  }\href {\doibase 10.1103/PhysRevD.62.023511} {\bibfield  {journal} {\bibinfo
  {journal} {\prd}\ }\textbf {\bibinfo {volume} {62}},\ \bibinfo {eid} {023511}
  (\bibinfo {year} {2000})},\ \Eprint {http://arxiv.org/abs/astro-ph/9912463}
  {astro-ph/9912463} \BibitemShut {NoStop}%
\bibitem [{\citenamefont {{Hamed}}\ \emph {et~al.}(2004)\citenamefont
  {{Hamed}}, \citenamefont {{Cheng}}, \citenamefont {{Luty}},\ and\
  \citenamefont {{Mukohyama}}}]{Hamed2004}%
  \BibitemOpen
  \bibfield  {author} {\bibinfo {author} {\bibfnamefont {N.~A.}\ \bibnamefont
  {{Hamed}}}, \bibinfo {author} {\bibfnamefont {H.~S.}\ \bibnamefont
  {{Cheng}}}, \bibinfo {author} {\bibfnamefont {M.~A.}\ \bibnamefont {{Luty}}},
  \ and\ \bibinfo {author} {\bibfnamefont {S.}~\bibnamefont {{Mukohyama}}},\
  }\href {\doibase 10.1088/1126-6708/2004/05/074} {\bibfield  {journal}
  {\bibinfo  {journal} {Journal of High Energy Physics}\ }\textbf {\bibinfo
  {volume} {5}},\ \bibinfo {eid} {074} (\bibinfo {year} {2004})},\ \Eprint
  {http://arxiv.org/abs/hep-th/0312099} {hep-th/0312099} \BibitemShut {NoStop}%
\bibitem [{\citenamefont {{Piazza}}\ and\ \citenamefont
  {{Tsujikawa}}(2004)}]{Piazza2004}%
  \BibitemOpen
  \bibfield  {author} {\bibinfo {author} {\bibfnamefont {F.}~\bibnamefont
  {{Piazza}}}\ and\ \bibinfo {author} {\bibfnamefont {S.}~\bibnamefont
  {{Tsujikawa}}},\ }\href {\doibase 10.1088/1475-7516/2004/07/004} {\bibfield
  {journal} {\bibinfo  {journal} {\jcap}\ }\textbf {\bibinfo {volume} {7}},\
  \bibinfo {eid} {004} (\bibinfo {year} {2004})},\ \Eprint
  {http://arxiv.org/abs/hep-th/0405054} {hep-th/0405054} \BibitemShut {NoStop}%
\bibitem [{\citenamefont {{Scherrer}}(2004)}]{Scherrer2004}%
  \BibitemOpen
  \bibfield  {author} {\bibinfo {author} {\bibfnamefont {R.~J.}\ \bibnamefont
  {{Scherrer}}},\ }\href {\doibase 10.1103/PhysRevLett.93.011301} {\bibfield
  {journal} {\bibinfo  {journal} {Physical Review Letters}\ }\textbf {\bibinfo
  {volume} {93}},\ \bibinfo {eid} {011301} (\bibinfo {year} {2004})},\ \Eprint
  {http://arxiv.org/abs/astro-ph/0402316} {astro-ph/0402316} \BibitemShut
  {NoStop}%
\bibitem [{\citenamefont {{Mukhanov}}\ and\ \citenamefont
  {{Vikman}}(2006)}]{Mukhanov2006}%
  \BibitemOpen
  \bibfield  {author} {\bibinfo {author} {\bibfnamefont {V.}~\bibnamefont
  {{Mukhanov}}}\ and\ \bibinfo {author} {\bibfnamefont {A.}~\bibnamefont
  {{Vikman}}},\ }\href {\doibase 10.1088/1475-7516/2006/02/004} {\bibfield
  {journal} {\bibinfo  {journal} {\jcap}\ }\textbf {\bibinfo {volume} {2}},\
  \bibinfo {eid} {004} (\bibinfo {year} {2006})},\ \Eprint
  {http://arxiv.org/abs/astro-ph/0512066} {astro-ph/0512066} \BibitemShut
  {NoStop}%
\bibitem [{\citenamefont {{Deffayet}}\ \emph {et~al.}(2010)\citenamefont
  {{Deffayet}}, \citenamefont {{Pujol{\`a}s}}, \citenamefont {{Sawicki}},\ and\
  \citenamefont {{Vikman}}}]{Deffayet2010}%
  \BibitemOpen
  \bibfield  {author} {\bibinfo {author} {\bibfnamefont {C.}~\bibnamefont
  {{Deffayet}}}, \bibinfo {author} {\bibfnamefont {O.}~\bibnamefont
  {{Pujol{\`a}s}}}, \bibinfo {author} {\bibfnamefont {I.}~\bibnamefont
  {{Sawicki}}}, \ and\ \bibinfo {author} {\bibfnamefont {A.}~\bibnamefont
  {{Vikman}}},\ }\href {\doibase 10.1088/1475-7516/2010/10/026} {\bibfield
  {journal} {\bibinfo  {journal} {\jcap}\ }\textbf {\bibinfo {volume} {10}},\
  \bibinfo {eid} {026} (\bibinfo {year} {2010})},\ \Eprint
  {http://arxiv.org/abs/1008.0048} {arXiv:1008.0048 [hep-th]} \BibitemShut
  {NoStop}%
\bibitem [{\citenamefont {{Pujol{\`a}s}}\ \emph {et~al.}(2011)\citenamefont
  {{Pujol{\`a}s}}, \citenamefont {{Sawicki}},\ and\ \citenamefont
  {{Vikman}}}]{Pujolas2011}%
  \BibitemOpen
  \bibfield  {author} {\bibinfo {author} {\bibfnamefont {O.}~\bibnamefont
  {{Pujol{\`a}s}}}, \bibinfo {author} {\bibfnamefont {I.}~\bibnamefont
  {{Sawicki}}}, \ and\ \bibinfo {author} {\bibfnamefont {A.}~\bibnamefont
  {{Vikman}}},\ }\href {\doibase 10.1007/JHEP11(2011)156} {\bibfield  {journal}
  {\bibinfo  {journal} {Journal of High Energy Physics}\ }\textbf {\bibinfo
  {volume} {11}},\ \bibinfo {eid} {156} (\bibinfo {year} {2011})},\ \Eprint
  {http://arxiv.org/abs/1103.5360} {arXiv:1103.5360 [hep-th]} \BibitemShut
  {NoStop}%
\bibitem [{\citenamefont {{Zlosnik}}\ \emph {et~al.}(2008)\citenamefont
  {{Zlosnik}}, \citenamefont {{Ferreira}},\ and\ \citenamefont
  {{Starkman}}}]{Zlosnik2008}%
  \BibitemOpen
  \bibfield  {author} {\bibinfo {author} {\bibfnamefont {T.~G.}\ \bibnamefont
  {{Zlosnik}}}, \bibinfo {author} {\bibfnamefont {P.~G.}\ \bibnamefont
  {{Ferreira}}}, \ and\ \bibinfo {author} {\bibfnamefont {G.~D.}\ \bibnamefont
  {{Starkman}}},\ }\href {\doibase 10.1103/PhysRevD.77.084010} {\bibfield
  {journal} {\bibinfo  {journal} {\prd}\ }\textbf {\bibinfo {volume} {77}},\
  \bibinfo {eid} {084010} (\bibinfo {year} {2008})},\ \Eprint
  {http://arxiv.org/abs/0711.0520} {arXiv:0711.0520} \BibitemShut {NoStop}%
\bibitem [{\citenamefont {{Battye}}\ \emph
  {et~al.}(2017{\natexlab{a}})\citenamefont {{Battye}}, \citenamefont
  {{Pace}},\ and\ \citenamefont {{Trinh}}}]{Battye2017}%
  \BibitemOpen
  \bibfield  {author} {\bibinfo {author} {\bibfnamefont {R.~A.}\ \bibnamefont
  {{Battye}}}, \bibinfo {author} {\bibfnamefont {F.}~\bibnamefont {{Pace}}}, \
  and\ \bibinfo {author} {\bibfnamefont {D.}~\bibnamefont {{Trinh}}},\ }\href
  {\doibase 10.1103/PhysRevD.96.064041} {\bibfield  {journal} {\bibinfo
  {journal} {\prd}\ }\textbf {\bibinfo {volume} {96}},\ \bibinfo {eid} {064041}
  (\bibinfo {year} {2017}{\natexlab{a}})},\ \Eprint
  {http://arxiv.org/abs/1707.06508} {arXiv:1707.06508} \BibitemShut {NoStop}%
\bibitem [{\citenamefont {{de Rham}}(2014)}]{deRham2014}%
  \BibitemOpen
  \bibfield  {author} {\bibinfo {author} {\bibfnamefont {C.}~\bibnamefont {{de
  Rham}}},\ }\href {\doibase 10.12942/lrr-2014-7} {\bibfield  {journal}
  {\bibinfo  {journal} {Living Reviews in Relativity}\ }\textbf {\bibinfo
  {volume} {17}},\ \bibinfo {eid} {7} (\bibinfo {year} {2014})},\ \Eprint
  {http://arxiv.org/abs/1401.4173} {arXiv:1401.4173 [hep-th]} \BibitemShut
  {NoStop}%
\bibitem [{\citenamefont {{G{\"u}mr{\"u}k{\c c}{\"u}o{\u g}lu}}\ \emph
  {et~al.}(2012)\citenamefont {{G{\"u}mr{\"u}k{\c c}{\"u}o{\u g}lu}},
  \citenamefont {{Kuroyanagi}}, \citenamefont {{Lin}}, \citenamefont
  {{Mukohyama}},\ and\ \citenamefont {{Tanahashi}}}]{Gumrukcuoglu2012}%
  \BibitemOpen
  \bibfield  {author} {\bibinfo {author} {\bibfnamefont {A.~E.}\ \bibnamefont
  {{G{\"u}mr{\"u}k{\c c}{\"u}o{\u g}lu}}}, \bibinfo {author} {\bibfnamefont
  {S.}~\bibnamefont {{Kuroyanagi}}}, \bibinfo {author} {\bibfnamefont
  {C.}~\bibnamefont {{Lin}}}, \bibinfo {author} {\bibfnamefont
  {S.}~\bibnamefont {{Mukohyama}}}, \ and\ \bibinfo {author} {\bibfnamefont
  {N.}~\bibnamefont {{Tanahashi}}},\ }\href {\doibase
  10.1088/0264-9381/29/23/235026} {\bibfield  {journal} {\bibinfo  {journal}
  {Classical and Quantum Gravity}\ }\textbf {\bibinfo {volume} {29}},\ \bibinfo
  {eid} {235026} (\bibinfo {year} {2012})},\ \Eprint
  {http://arxiv.org/abs/1208.5975} {arXiv:1208.5975 [hep-th]} \BibitemShut
  {NoStop}%
\bibitem [{\citenamefont {{G{\"u}mr{\"u}k{\c c}{\"u}o{\u g}lu}}\ \emph
  {et~al.}(2013)\citenamefont {{G{\"u}mr{\"u}k{\c c}{\"u}o{\u g}lu}},
  \citenamefont {{Hinterbichler}}, \citenamefont {{Lin}}, \citenamefont
  {{Mukohyama}},\ and\ \citenamefont {{Trodden}}}]{Gumrukcuoglu2013}%
  \BibitemOpen
  \bibfield  {author} {\bibinfo {author} {\bibfnamefont {A.~E.}\ \bibnamefont
  {{G{\"u}mr{\"u}k{\c c}{\"u}o{\u g}lu}}}, \bibinfo {author} {\bibfnamefont
  {K.}~\bibnamefont {{Hinterbichler}}}, \bibinfo {author} {\bibfnamefont
  {C.}~\bibnamefont {{Lin}}}, \bibinfo {author} {\bibfnamefont
  {S.}~\bibnamefont {{Mukohyama}}}, \ and\ \bibinfo {author} {\bibfnamefont
  {M.}~\bibnamefont {{Trodden}}},\ }\href {\doibase 10.1103/PhysRevD.88.024023}
  {\bibfield  {journal} {\bibinfo  {journal} {\prd}\ }\textbf {\bibinfo
  {volume} {88}},\ \bibinfo {eid} {024023} (\bibinfo {year} {2013})},\ \Eprint
  {http://arxiv.org/abs/1304.0449} {arXiv:1304.0449 [hep-th]} \BibitemShut
  {NoStop}%
\bibitem [{\citenamefont {{Battye}}\ and\ \citenamefont
  {{Moss}}(2007)}]{Battye2007}%
  \BibitemOpen
  \bibfield  {author} {\bibinfo {author} {\bibfnamefont {R.~A.}\ \bibnamefont
  {{Battye}}}\ and\ \bibinfo {author} {\bibfnamefont {A.}~\bibnamefont
  {{Moss}}},\ }\href {\doibase 10.1103/PhysRevD.76.023005} {\bibfield
  {journal} {\bibinfo  {journal} {\prd}\ }\textbf {\bibinfo {volume} {76}},\
  \bibinfo {eid} {023005} (\bibinfo {year} {2007})},\ \Eprint
  {http://arxiv.org/abs/astro-ph/0703744} {astro-ph/0703744} \BibitemShut
  {NoStop}%
\bibitem [{\citenamefont {{Pearson}}(2014)}]{Pearson2014}%
  \BibitemOpen
  \bibfield  {author} {\bibinfo {author} {\bibfnamefont {J.~A.}\ \bibnamefont
  {{Pearson}}},\ }\href {\doibase 10.1002/andp.201400052} {\bibfield  {journal}
  {\bibinfo  {journal} {Annalen der Physik}\ }\textbf {\bibinfo {volume}
  {526}},\ \bibinfo {pages} {318} (\bibinfo {year} {2014})},\ \Eprint
  {http://arxiv.org/abs/1403.1213} {arXiv:1403.1213} \BibitemShut {NoStop}%
\bibitem [{\citenamefont {{Battye}}\ \emph {et~al.}(2016)\citenamefont
  {{Battye}}, \citenamefont {{Bolliet}},\ and\ \citenamefont
  {{Pearson}}}]{Battye2016a}%
  \BibitemOpen
  \bibfield  {author} {\bibinfo {author} {\bibfnamefont {R.~A.}\ \bibnamefont
  {{Battye}}}, \bibinfo {author} {\bibfnamefont {B.}~\bibnamefont {{Bolliet}}},
  \ and\ \bibinfo {author} {\bibfnamefont {J.~A.}\ \bibnamefont {{Pearson}}},\
  }\href {\doibase 10.1103/PhysRevD.93.044026} {\bibfield  {journal} {\bibinfo
  {journal} {\prd}\ }\textbf {\bibinfo {volume} {93}},\ \bibinfo {eid} {044026}
  (\bibinfo {year} {2016})},\ \Eprint {http://arxiv.org/abs/1508.04569}
  {arXiv:1508.04569} \BibitemShut {NoStop}%
\bibitem [{\citenamefont {{Cheung}}\ \emph {et~al.}(2008)\citenamefont
  {{Cheung}}, \citenamefont {{Fitzpatrick}}, \citenamefont {{Kaplan}},
  \citenamefont {{Senatore}},\ and\ \citenamefont {{Creminelli}}}]{Cheung2008}%
  \BibitemOpen
  \bibfield  {author} {\bibinfo {author} {\bibfnamefont {C.}~\bibnamefont
  {{Cheung}}}, \bibinfo {author} {\bibfnamefont {A.~L.}\ \bibnamefont
  {{Fitzpatrick}}}, \bibinfo {author} {\bibfnamefont {J.}~\bibnamefont
  {{Kaplan}}}, \bibinfo {author} {\bibfnamefont {L.}~\bibnamefont
  {{Senatore}}}, \ and\ \bibinfo {author} {\bibfnamefont {P.}~\bibnamefont
  {{Creminelli}}},\ }\href {\doibase 10.1088/1126-6708/2008/03/014} {\bibfield
  {journal} {\bibinfo  {journal} {Journal of High Energy Physics}\ }\textbf
  {\bibinfo {volume} {3}},\ \bibinfo {eid} {014-014} (\bibinfo {year}
  {2008})},\ \Eprint {http://arxiv.org/abs/0709.0293} {arXiv:0709.0293
  [hep-th]} \BibitemShut {NoStop}%
\bibitem [{\citenamefont {{Creminelli}}\ \emph {et~al.}(2009)\citenamefont
  {{Creminelli}}, \citenamefont {{D'Amico}}, \citenamefont {{Nore{\~n}a}},\
  and\ \citenamefont {{Vernizzi}}}]{Creminelli2009}%
  \BibitemOpen
  \bibfield  {author} {\bibinfo {author} {\bibfnamefont {P.}~\bibnamefont
  {{Creminelli}}}, \bibinfo {author} {\bibfnamefont {G.}~\bibnamefont
  {{D'Amico}}}, \bibinfo {author} {\bibfnamefont {J.}~\bibnamefont
  {{Nore{\~n}a}}}, \ and\ \bibinfo {author} {\bibfnamefont {F.}~\bibnamefont
  {{Vernizzi}}},\ }\href {\doibase 10.1088/1475-7516/2009/02/018} {\bibfield
  {journal} {\bibinfo  {journal} {Journal of Cosmology and Astro-Particle
  Physics}\ }\textbf {\bibinfo {volume} {2}},\ \bibinfo {pages} {18} (\bibinfo
  {year} {2009})},\ \Eprint {http://arxiv.org/abs/0811.0827} {arXiv:0811.0827}
  \BibitemShut {NoStop}%
\bibitem [{\citenamefont {{Battye}}\ and\ \citenamefont
  {{Pearson}}(2012)}]{Battye2012}%
  \BibitemOpen
  \bibfield  {author} {\bibinfo {author} {\bibfnamefont {R.~A.}\ \bibnamefont
  {{Battye}}}\ and\ \bibinfo {author} {\bibfnamefont {J.~A.}\ \bibnamefont
  {{Pearson}}},\ }\href {\doibase 10.1088/1475-7516/2012/07/019} {\bibfield
  {journal} {\bibinfo  {journal} {\jcap}\ }\textbf {\bibinfo {volume} {7}},\
  \bibinfo {eid} {019} (\bibinfo {year} {2012})},\ \Eprint
  {http://arxiv.org/abs/1203.0398} {arXiv:1203.0398 [hep-th]} \BibitemShut
  {NoStop}%
\bibitem [{\citenamefont {{Battye}}\ and\ \citenamefont
  {{Pearson}}(2013{\natexlab{b}})}]{Battye2013a}%
  \BibitemOpen
  \bibfield  {author} {\bibinfo {author} {\bibfnamefont {R.~A.}\ \bibnamefont
  {{Battye}}}\ and\ \bibinfo {author} {\bibfnamefont {J.~A.}\ \bibnamefont
  {{Pearson}}},\ }\href {\doibase 10.1103/PhysRevD.88.084004} {\bibfield
  {journal} {\bibinfo  {journal} {\prd}\ }\textbf {\bibinfo {volume} {88}},\
  \bibinfo {eid} {084004} (\bibinfo {year} {2013}{\natexlab{b}})},\ \Eprint
  {http://arxiv.org/abs/1301.5042} {arXiv:1301.5042 [astro-ph.CO]} \BibitemShut
  {NoStop}%
\bibitem [{\citenamefont {{Battye}}\ \emph
  {et~al.}(2017{\natexlab{b}})\citenamefont {{Battye}}, \citenamefont
  {{Bolliet}},\ and\ \citenamefont {{Pace}}}]{Battye2017a}%
  \BibitemOpen
  \bibfield  {author} {\bibinfo {author} {\bibfnamefont {R.~A.}\ \bibnamefont
  {{Battye}}}, \bibinfo {author} {\bibfnamefont {B.}~\bibnamefont {{Bolliet}}},
  \ and\ \bibinfo {author} {\bibfnamefont {F.}~\bibnamefont {{Pace}}},\
  }\href@noop {} {\bibfield  {journal} {\bibinfo  {journal} {ArXiv e-prints}\ }
  (\bibinfo {year} {2017}{\natexlab{b}})},\ \Eprint
  {http://arxiv.org/abs/1712.05976} {arXiv:1712.05976} \BibitemShut {NoStop}%
\bibitem [{\citenamefont {{Ballesteros}}\ and\ \citenamefont
  {{Lesgourgues}}(2010)}]{Ballesteros2010}%
  \BibitemOpen
  \bibfield  {author} {\bibinfo {author} {\bibfnamefont {G.}~\bibnamefont
  {{Ballesteros}}}\ and\ \bibinfo {author} {\bibfnamefont {J.}~\bibnamefont
  {{Lesgourgues}}},\ }\href {\doibase 10.1088/1475-7516/2010/10/014} {\bibfield
   {journal} {\bibinfo  {journal} {\jcap}\ }\textbf {\bibinfo {volume} {10}},\
  \bibinfo {eid} {014} (\bibinfo {year} {2010})},\ \Eprint
  {http://arxiv.org/abs/1004.5509} {arXiv:1004.5509 [astro-ph.CO]} \BibitemShut
  {NoStop}%
\bibitem [{\citenamefont {{De Felice}}\ and\ \citenamefont
  {{Tsujikawa}}(2010)}]{DeFelice2010}%
  \BibitemOpen
  \bibfield  {author} {\bibinfo {author} {\bibfnamefont {A.}~\bibnamefont {{De
  Felice}}}\ and\ \bibinfo {author} {\bibfnamefont {S.}~\bibnamefont
  {{Tsujikawa}}},\ }\href {\doibase 10.12942/lrr-2010-3} {\bibfield  {journal}
  {\bibinfo  {journal} {Living Reviews in Relativity}\ }\textbf {\bibinfo
  {volume} {13}},\ \bibinfo {eid} {3} (\bibinfo {year} {2010})},\ \Eprint
  {http://arxiv.org/abs/1002.4928} {arXiv:1002.4928 [gr-qc]} \BibitemShut
  {NoStop}%
\bibitem [{\citenamefont {{Brax}}\ \emph {et~al.}(2008)\citenamefont {{Brax}},
  \citenamefont {{van de Bruck}}, \citenamefont {{Davis}},\ and\ \citenamefont
  {{Shaw}}}]{Brax2008}%
  \BibitemOpen
  \bibfield  {author} {\bibinfo {author} {\bibfnamefont {P.}~\bibnamefont
  {{Brax}}}, \bibinfo {author} {\bibfnamefont {C.}~\bibnamefont {{van de
  Bruck}}}, \bibinfo {author} {\bibfnamefont {A.-C.}\ \bibnamefont {{Davis}}},
  \ and\ \bibinfo {author} {\bibfnamefont {D.~J.}\ \bibnamefont {{Shaw}}},\
  }\href {\doibase 10.1103/PhysRevD.78.104021} {\bibfield  {journal} {\bibinfo
  {journal} {\prd}\ }\textbf {\bibinfo {volume} {78}},\ \bibinfo {eid} {104021}
  (\bibinfo {year} {2008})},\ \Eprint {http://arxiv.org/abs/0806.3415}
  {arXiv:0806.3415} \BibitemShut {NoStop}%
\end{thebibliography}%
